\documentclass[aps,prd,twocolumn,showpacs,superscriptaddress,preprintnumbers,dblfloatfix,floatfix,nofootinbib]{revtex4-2}
\usepackage[utf8]{inputenc}
\usepackage{amsmath}
\usepackage{graphicx}
\usepackage{caption}
\usepackage{amssymb}
\usepackage{color}
\usepackage{cancel}
\usepackage{framed}
\usepackage{comment}
\usepackage{mathtools}
\usepackage{hyperref}
\usepackage{chngcntr}
\usepackage[normalem]{ulem}

\counterwithin{equation}{section}

\newcommand{\vev}[1]{\langle #1 \rangle}

\newcommand{\nn}{\nonumber}

\def\a{\alpha} \def\b{\beta} \def\g{\gamma} 
\def\d{\delta} \def\e{\epsilon} 
 \def\th{\theta} 
\def\k{\kappa} \def\l{\lambda} \def\m{\mu} 
\def\n{\nu}  \def\x{\xi}   
 \def\s{\sigma}

 \def\D{\Delta}  
\def\L{\Lambda}

 \newcommand{\Lcal}{{\mathcal L}} 
 \newcommand{\Ncal}{{\mathcal N}}
\newcommand{\Ocal}{{\mathcal O}}

\def\bea  {\begin{eqnarray}}   \def\eea  {\end{eqnarray}}
\def\TeV{\,{\rm TeV}}
\def\GeV{\,{\rm GeV}}
\def\MeV{\,{\rm MeV}}
\def\keV{\,{\rm keV}}
\def\eV{\,{\rm eV}}
\def\Mpl{ M_{\rm Pl}}

\def\Weizmann{\small{Department of Particle Physics and Astrophysics, Weizmann Institute of Science, Rehovot 761001, Israel}}
\def\IPMU{\small{Kavli IPMU (WPI), UTIAS, The University of Tokyo, Kashiwa, Chiba 277-8583, Japan}}

\usepackage{tikz,xcolor,hyperref}
\definecolor{lime}{HTML}{A6CE39}
\DeclareRobustCommand{\orcidicon}{%
	\begin{tikzpicture}
	\draw[lime, fill=lime] (0,0) 
	circle [radius=0.16] 
	node[white] {{\fontfamily{qag}\selectfont \tiny ID}};	\draw[white, fill=white] (-0.0625,0.095) 
	circle [radius=0.007];	\end{tikzpicture}
	\hspace{-2mm}}
\foreach \x in {A,...,Z}{%
	\expandafter\xdef\csname orcid\x\endcsname{\noexpand\href{https://orcid.org/\csname orcidauthor\x\endcsname}{\noexpand\orcidicon}}
}

\begin{document}

\preprint{\hfill IPMU22-0051}

\date{\today}
\title{\Large\bfseries From axion quality and naturalness problems to a \\ high-quality $\mathbb{Z}_{N}$ QCD relaxion}

\author{Abhishek Banerjee\orcidA{}\,}
\email{abhishek.banerjee@weizmann.ac.il }
\affiliation{\Weizmann}

\author{Joshua Eby\orcidB{}\,}
\email{joshaeby@gmail.com}
\affiliation{\IPMU}

\author{Gilad Perez\orcidC{}\, }
\email{gilad.perez@weizmann.ac.il }
\affiliation{\Weizmann}

\begin{abstract}
We highlight general issues associated with quality and naturalness problems in theories of light QCD-axions, axion-like particles, and relaxions. We show the presence of Planck-suppressed operators generically lead to scalar coupling of axions with the Standard model. 
We present a new class of $\mathbb Z_{N}$ QCD relaxion models that can address both the QCD relaxion CP problem as well as its quality problem. 
This new class of models also leads to interesting experimental signatures, which can be searched for at the precision frontier.
\end{abstract}

\maketitle

\section{Introduction}

The Standard Model of particle physics (SM) is an extremely successful yet incomplete description of nature. 
It cannot account for the observed neutrino masses and mixings, the matter anti-matter asymmetry, and the origin of Dark Matter (DM). Even within the framework of the SM, we have the Higgs-hierarchy and the Strong CP problems. 
On top of that, the effect of gravity is expected to be significant at the Planck scale despite the lack of knowledge about its quantum nature.
In particular,
quantum gravity is expected to violate global symmetries in the UV, implying the existence of symmetry-breaking operators suppressed by powers of the Planck mass $\Mpl=2.4\times10^{18}\GeV$ in the framework of effective field theory (EFT). 
For an axion field $\Phi$ with a global Peccei-Quinn (PQ) symmetry \cite{Peccei:1977hh}, one for instance expects, among others, operators of the form
\begin{equation} \label{eq:PQB1}
 \Lcal \supset \frac{1}{2}\left(\frac{c_N\Phi^N + {\rm h.c.}}{\Mpl^{N}}\right) \Ocal
\end{equation}
where $N$ is an integer, $c_N$ is a dimensionless EFT parameter, and $\Ocal$ is any dimension-four operator consistent with the unbroken gauge symmetries. Expanding $\Phi = f\,e^{i\,\phi/f}$, this Lagrangian generates a shift-symmetric potential of the form
\begin{align} \label{eq:PQB2}
 V_\D &= |c_N|\D^N \cos\left(\frac{N\,\phi}{f} + \b\right) \Ocal,
\end{align}
where $\D\equiv f/\Mpl$ and $\b=\arg(c_N)$ is an arbitrary phase which is generically $\Ocal(1)$. Note that, if CP is not broken by gravity then $\b = 0$. 
The dimension $N$ of the PQ-breaking operator in Eqs. (\ref{eq:PQB1}-\ref{eq:PQB2}) is dictated by the unbroken gauge symmetries of the underlying theory.

The leading contribution to PQ-breaking arises from a constant operator multiplied by $\Mpl^4$ to match the dimension. 
This definition fixes $N>4$ so that these operators are suppressed in the limit $\Mpl \to \infty$ (see e.g. \cite{Hook:2018dlk,DiLuzio:2021pxd}).
This implies a contribution $V_\D = |c_N| \D^N \Mpl^4 \cos\left(N\phi/f + \b\right)$ to the scalar field theory.
If this field is identified with the QCD axion~\cite{Weinberg:1977ma,Wilczek:1977pj,Dine:1981rt,Zhitnitsky:1980tq,Kim:1979if,Shifman:1979if}, then the coefficient $|c_N|\D^N$ in Eq.~\eqref{eq:PQB2} cannot be too large or else it will spoil the solution to the strong CP problem; this is the so-called \emph{axion quality problem}~\cite{Kamionkowski:1992mf,Barr:1992qq,Ghigna:1992iv}, and it can be solved by either (a) fine-tuning, e.g. taking $|c_N| \ll 1$, (b) taking $f$ very small (which is constrained by measurements of axion couplings to matter), or (c) forbidding operators of dimension $N$ up to some large value, for example by imposing some unbroken gauge symmetry (e.g. $\mathbb{Z}_N$).

We describe the constraints on these operators in greater detail in the subsections below.

\section{Axion phenomenology}
\subsection{Axion-like particles and naturalness} \label{sec:ALP}

A general axion-like particle (ALP) which is not coupled to QCD does not exhibit a quality problem related to the vacuum structure (see next section), and therefore it might seem that the presence of Planck-suppressed operators would be harmless. However, these same operators can induce large contributions to the ALP mass, leading to a fine-tuning problem. 

Planck-suppressed operators can also generate ALP couplings to the SM scalar operators, which we discuss in Section \ref{sec:CPV}. 
In the absence of any CP violation, ALPs interact with the SM scalar operators quadratically at the leading order, whereas if gravity does not respect CP, i.e. for $\beta\neq0$ in Eq.~\eqref{eq:PQB2}, these interactions are generated at linear order.

An ALP is defined by its mass $m$ and coupling with the SM pseudoscalar operators. These couplings are associated with an energy scale, which we will identify with $f$. 
To analyse the effect of Planck-suppressed operators, we consider an ALP potential of 
\begin{equation}
 V_{\rm ALP}(\phi) = -m^2f^2\cos\left(\frac{\phi}{f}\right),
\end{equation}
which defines the ALP mass $m$. 
However, the second derivative of the potential induced by Planck-suppressed operators in Eq.~\eqref{eq:PQB2} is
\begin{align}
 V_\D''(\phi) 
 		&= |c_N|\D^{N-2} \Mpl^2\,N^2\cos\left(\frac{N\phi}{f} + \b\right). 
\end{align} 
Therefore, at leading order in $\phi/f \ll 1$, we have a bare contribution to the mass $m^2$ and a correction of order
\begin{equation}
 \d m^2 \simeq |c_N|\cos\b\,\D^{N-2} N^2 \Mpl^2\,.
\end{equation}
Such corrections satisfy $\d m^2 \ll m^2$ only if
\begin{align} \label{eq:epALP}
 &\left|\frac{|c_N|\cos\b\,\D^{N-2}N^2 \Mpl^2}{m^2}\right| 
 				= \left|\frac{|c_N|\cos\b\,\D^{N}N^2 \Mpl^4}{m^2f^2}\right|  \ll 1\,.
\end{align}

Assuming $c_N \sim \cos\b \sim 1$, one can translate the inequality~\eqref{eq:epALP} into an upper bound on $f$ as a function of $N$ in order to have negligible fine-tuning of the ALP mass. We illustrate these limits for ALP masses $m=1,10^{-7},10^{-14}$ eV using the red, blue, and green dotted lines (respectively) in Figure \ref{fig:fmax}.

\subsection{QCD axion quality and naturalness} \label{sec:QCD}

QCD axions \cite{Weinberg:1977ma,Wilczek:1977pj,Dine:1981rt,Zhitnitsky:1980tq,Kim:1979if,Shifman:1979if} exhibit a quality problem when the contribution of Planck-suppressed operators contribute significantly to a shift in the low-energy vacuum of the potential \cite{Kamionkowski:1992mf,Barr:1992qq,Ghigna:1992iv}. At low energy, QCD axions have a potential of the form \cite{DiVecchia:1980yfw,GrillidiCortona:2015jxo}
\begin{align} \label{eq:VQCD}
 V_a(\phi) = &-\L_{\rm QCD}^3 (m_u+m_d) \nn \\
 		&\times \sqrt{1 - \frac{2z}{(1+z)^2}\left[1 - \cos\left(\frac{\phi}{f} + \bar{\th}\right)\right]},
\end{align}
where $z = m_u/m_d \simeq 0.485$~\cite{Fodor:2016bgu,Workman:2022ynf} is the ratio of up and down quark masses, $\L_{\rm QCD} = \vev{\bar{q}q}^{1/3}$ is the QCD scale defined by the quark condensate, and $\bar{\th}$ is the effective CP violating angle. 
At leading order in $z\ll 1$ (and ignoring an irrelevant constant), we have
\begin{equation}
 V_a(\phi) \simeq  - \L_{a}^4 \cos\left(\frac{\phi}{f} + \bar\theta\right),\label{eq:QCD_axion_simplified}
\end{equation}
where for simplicity we define $\L_a = (\L_{\rm QCD}^3 m_u)^{1/4} 
\simeq 84\MeV$.

In the presence of the leading Planck-suppressed operator, one can find the minimum of the QCD-axion potential as
\begin{align}
 0 &= V'(\vev{\phi}) \nn \\
 				&= |c_N|\D^N N\,\Mpl^4 \sin\left(N\e + \b'\right) + \L_{a}^4 \sin\e \nn \\
 				&\approx |c_N|\D^N N\,\Mpl^4 \sin\b' + \L_{a}^4 \e,
\end{align}
where $\e \equiv \vev{\phi}/f - \bar\th$ and $\b' \equiv \b - N\bar\th$ which is generically $\Ocal(1)$. 
Non-observation of the neutron electric dipole moment (EDM) implies that $|\e| \lesssim 10^{-10}$ (see e.g. \cite{Alexandrou:2020mds,Pignol:2021uuy}), so in the last step we have expanded in small $\e,\,N\e \ll 1$. 
In order to not spoil the QCD axion solution to the strong CP problem, one must require
\begin{equation} \label{eq:epQCD}
 |\e| = \left|\frac{|c_N| \sin\b'\,\D^N \,N\Mpl^4}{\L_{a}^4}\right| \lesssim 10^{-10}.
\end{equation}
At leading order in $N$ this gives (for $c_N\simeq \sin\b'\simeq 1$)
\begin{equation}
 N \gtrsim  \frac{\log\left(10^{-10}(\L_{a}/\Mpl)^4\right)}{\log\left(\D\right)} 
 				= \frac{201}{19-\log\left(f/10^{10}\,{\rm GeV}\right)},
\end{equation}
see also \cite{Holman:1992us}.
So for PQ quality to be preserved, one needs to forbid operators with $N\lesssim 10$ ($13$) for $f=10^{10}$ ($10^{12}$) GeV. A simple way to do this is with a gauged $\mathbb{Z}_N$ symmetry (see Section \ref{sec:ZN}). 

\begin{figure}[t]
 \centering
 \includegraphics[scale=.85]{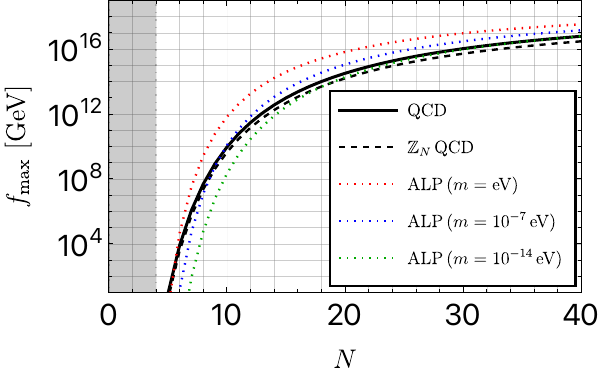}
 \caption{The value of $f_{\rm max}$ for a given operator dimension $N$ which satisfies the quality problem constraint for ordinary QCD (Eq. \eqref{eq:epQCD}, black solid) or $\mathbb{Z}_N$ QCD (Eq. \eqref{eq:epZN}, black dashed), compared to $f_{\rm max}$ to satisfy $\d m/m \ll 1$ in ALP case for $m=1,10^{-7},10^{-14}$ eV (Eq. \eqref{eq:epALP}, red, blue, and green dotted curves, respectively).}
 \label{fig:fmax}
\end{figure}

The inequality of \eqref{eq:epQCD} is illustrated by the black solid line in Figure \ref{fig:fmax}. Comparing the QCD case to an ALP where $m^2f^2 \simeq \L_{a}^4$, we observe a natural suppression of $10^{-10}N$ in the ALP naturalness condition in Eq.~\eqref{eq:epALP}, relative to Eq. \eqref{eq:epQCD}. Further, ALPs can populate a wider space of values for $m$ and $f$, allowing for more freedom in parameter inputs. Still, it is intriguing that the requirement of natural ALP mass given in \eqref{eq:epALP} is nearly as restrictive as the quality problem for QCD axions. 

As we point out above, generically scalar fields acquire large mass corrections from Planck-suppressed operators. Therefore in principle there is another constraint on the quality of the QCD axion, arising from fine-tuning of the axion mass, though this is always weaker than the constraint above (this was also pointed out in \cite{Holman:1992us}). Finally, note that in principle one could satisfy Eq. \eqref{eq:epQCD} even at small $N$ by tuning the EFT coefficient $|c_N|\ll 1$ or the phase parameter $|\b'| = |\b - N\bar{\th}| \ll 1$. However, this quickly leads to a fine-tuning as bad as (or worse than) the original strong CP problem.

\subsection{High-quality, natural $\mathbb{Z}_N$ QCD Axion} \label{sec:ZN}

It was shown in \cite{Hook:2018jle} that an extended sector with $N$ copies of the SM, related by a $\mathbb{Z}_N$ symmetry, can lead to a QCD-like axion of mass much smaller than that of canonical QCD axion, due to additional suppression by $\sim z^N$ in the effective QCD scale. This idea was further investigated in \cite{DiLuzio:2021pxd} and shown to simultaneously admit a viable ultralight axion DM candidate \cite{DiLuzio:2021gos}. If this $\mathbb{Z}_N$ symmetry is gauged, it can protect the theory from Planck-suppressed operators in Eq.~\eqref{eq:PQB1}.

Let us consider $N$ copies of the SM which are related to each other by a $\mathbb{Z}_N$ symmetry which is non-linearly realized by the axion field $\phi$, as 
\bea
\mathbb{Z}_N: \,\, &&{\rm SM}_k\to {\rm SM}_{k+1({\rm mod}\, N)}\\
&& \quad\,\,  \phi \to \phi + \frac{2 \pi k}{N}f\,,
\eea  
with $k=0,\cdots,N-1$. 
The most general $\mathbb{Z}_N$ symmetric Lagrangian\footnote{Note that there could also be portal couplings between sectors, though we postpone discussion of this to Section \ref{sec:relaxedpheno}.} can be written as
\bea
\!\!\!\!\!
\mathcal L = \sum_{k=0}^{N-1} \mathcal{L}_{{\rm SM}_k}+ \frac{\alpha_s}{8\pi}\left(\frac{\phi}{f}+\bar\theta+\frac{2\pi k}{N}\right) G_k\tilde G_k\,.
\eea
The axion will receive contributions from all the $N$ sectors; the combined potential can be written as
\bea \label{eq:VtotRR}
V_{\rm tot}(\phi) = \sum_{k=0}^{N-1} V\left(\frac{\phi}{f} + \bar{\th} + \frac{2\pi k}{N}\right)\,,
\eea
where, the axion potential in each sector is
\bea 
V(x) &=& - \Lambda_{\rm QCD}^3 m_u \sqrt{1+ z^2+2z \cos x}\,, \nn
\eea
as shown in Eq.~\eqref{eq:VQCD}.  

At low energies, this theory differs from the generic QCD case because the effective QCD scale is shifted. 
This is apparent in the effective potential of the theory \cite{DiLuzio:2021pxd} (see Eq. (2.30)):
\begin{equation} \label{eq:VN}
 V_N(\phi) \simeq -\sqrt{\frac{1-z^2}{\pi\,N}}(-z)^{N-1} \L_{\rm QCD}^3 m_u
 						\cos\left(\frac{N\,\phi}{f} + \bar{\th}\right).
\end{equation}
The requirement $V'(\phi) = V_\D'(\phi) + V_N'(\phi) = 0$ implies
\begin{equation} \label{eq:epZN}
 |\e| = \left|\frac{|c_N| \sin\b'\,\D^N\,\Mpl^4}{\L_{a}^4}
 \frac{1}{\k}\right| \lesssim 10^{-10},
\end{equation}
where $\k\equiv z^{N-1}\sqrt{(1-z^2)/(\pi\,N)}$. 

The $\mathbb{Z}_N$ axion case of Refs. \cite{Hook:2018jle,DiLuzio:2021pxd,DiLuzio:2021gos} is illustrated by the black dashed line in Figure \ref{fig:fmax}. 
The symmetry provides a mechanism for suppressing operators up to some large $N$ relative to the vanilla QCD case; however, at any given $N$, the inequality ~\eqref{eq:epZN} has a natural enhancement of order $1/\k \gg 1$ relative to the minimal QCD axion (c.f. Eq. \eqref{eq:epQCD}).

\subsection{Challenges associated with the QCD relaxion idea}
\label{sec:challenges}

The relaxion framework, proposed in~\cite{Graham:2015cka} provides a new insight on the hierarchy problem, which does not require TeV-scale new physics, but rather implies a non-trivial cosmological evolution of the Higgs mass. 
The original relaxion model was based on the QCD axion model~\cite{Graham:2015cka}\footnote{See~\cite{Espinosa:2015eda} this for a possible generalisation of the back-reaction potential, and ~\cite{Hook:2016mqo,Fonseca:2018xzp,Fonseca:2019lmc} for non-inflationary relaxation mechanism.}. 

However, as the back-reaction and the rolling potential are sequestered, the relaxion stopping point corresponds to sizeable phase, and generically cannot be set to zero. 
It was noticed in the original paper~\cite{Graham:2015cka} as well. 
Furthermore, as was shown in~\cite{Banerjee:2020kww}, and further derived below for the QCD-relaxion model, the peculiar nature of the relaxion dynamics implies that the relaxion stops at a highly non-generic point in the field space. 
At this point, the mass is parametrically suppressed, and the phase
is predicted to be very close to $\pi/2$, 
a mechanism dubbed the \textit{relaxed relaxion}. 
In~\cite{Nelson:2017cfv} a solution was proposed to this problem; however, it required non-classical evolution of the relaxion and thus, led to further problems associated with the measure problem~\cite{Gupta:2018wif}. 

In addition to that, a successful relaxation of the Higgs mass requires large hierarchy between the scales of the rolling potential and the back-reaction potential~\cite{Gupta:2015uea} and thus, the relaxion setup rely on a carefully designed potential derived from the clockwork mechanism~\cite{Choi:2014rja,Choi:2015fiu,Kaplan:2015fuy,Giudice:2016yja}, which is based on a $U(1)$ global symmetry. 
The resulting construction suffers from a fairly severe quality problem, unless the relaxion is rather heavy~\cite{Davidi:2017gir}. 
In Section~\ref{sec:HQQCD}, we propose a new construction that addresses both of the above challenges.

\subsection{Axion/ALP couplings from unknown Planck physics} \label{sec:CPV}

As mentioned previously, the Planck-suppressed PQ-breaking operators in Eq.~\eqref{eq:PQB2} give rise to SM couplings. 
This is, as we discuss below, due to the fact that the additional terms 
may be misaligned in phase relative to the terms induced by the IR QCD instantons. 
In the presence of CP violation, the resulting couplings can be linear in the field, whereas if CP is conserved the leading couplings are quadratic.
In addition to that, the QCD axion always induces a scalar interaction with the nucleons at the quadratic order of the axion field~\cite{Kim:2022ype}. 

For low-energy phenomenology, we consider ALP/axion interaction with the electrons, photons, or gluons; the Lagrangian of such interactions can be written as
\begin{align}\label{eq:Lscalar}
 \Lcal &\supset \frac{\phi}{\Mpl}\left[d_{m_e}^{(1)} m_e\bar{e}e 
 			+ \frac{d_\a^{(1)}}{4}F^2 + \frac{d_{g}^{(1)}\beta(g)}{2g} G^2\right] \nn \\
		& + \frac{\phi^2}{2\Mpl^2}
		 \left[d_{m_e}^{(2)} m_e\bar{e}e 
 			+ \frac{d_\a^{(2)}}{4}F^2 + \frac{d_{g}^{(2)}\beta(g)}{2g} G^2\right].
\end{align}
where, $e$ is the electron field, $F^2=F^{\mu\nu}F_{\mu\nu}$, $G^2=\frac{1}{2}{\rm Tr}(G^{\mu\nu}G_{\mu\nu})$, $F_{\mu\nu}$ ($G_{\mu\nu}$) is the electromagnetic (QCD) field strength. Also, $g$ is the QCD gauge coupling and $\beta(g)$ is the beta function. 
Such couplings can be searched for via the equivalence principle violations and/or fifth forces experiments~\cite{Damour:2010rp,Smith:1999cr,Schlamminger:2007ht,Berge:2017ovy,Touboul:2017grn,Touboul:2022yrw,Hees:2018fpg}, or oscillation of fundamental constants (for a review, see for example~\cite{Safronova:2017xyt}; for proposals, see~\cite{Arvanitaki:2014faa,Stadnik:2014tta,Stadnik:2015xbn,Arvanitaki:2015iga,Safronova:2018quw,Dzuba:2018fed,Geraci:2018fax,Manley:2019vxy,Grote:2019uvn,Banerjee:2020kww,Peik:2020cwm}; and for experiments providing bounds on oscillations see~\cite{VanTilburg:2015oza, Aharony:2019iad, Hees:2016gop,Antypas:2019qji, Kennedy:2020bac, Wcislo:2018ojh,Vermeulen:2021epa, Campbell:2020fvq, Savalle:2020vgz,Aiello:2021wlp,Oswald:2021vtc,Tretiak:2022ndx}). Note that, one can also consider ALP/axion interaction with $m_q\bar q q$, where $q=u,d$ denotes the light quarks; see e.g.~\cite{Oswald:2021vtc} for bounds on such couplings.

To see how the above interactions are generated from Eq. \eqref{eq:PQB2}, one can expand the cosine part up to quadratic order to find
\begin{equation} \label{eq:cosexp}
 \cos\left(\frac{N\,\phi}{f} + \b\right) = \cos\b - \sin\b\,\frac{N\phi}{f} - \frac{\cos\b}{2}\left(\frac{N\phi}{f}\right)^2 +\cdots .
\end{equation} 
Comparing Eqs.~\eqref{eq:PQB2} and \eqref{eq:cosexp}, we can easily identify 
\begin{equation} \label{eq:dX}
 d_{X}^{(1)} = |c_N| N\, \sin\b\,\D^{N-1}, \qquad d_{X}^{(2)} = |c_N| N^2 \cos\b\,\D^{N-2},
\end{equation}
for $X=m_e,\a,g$, which we will refer to as the \emph{quality couplings} of the theory (due to their possible connection with the quality problem). As discussed before, if gravity respects CP, then $\beta=0$ and thus, there is no linear scalar coupling between ALP and SM. However, the quadratic interactions are present both for the CP-violating and CP-conserving cases.

Experimental searches for equivalence principle violations and fifth forces \cite{Smith:1999cr,Schlamminger:2007ht,Berge:2017ovy,Touboul:2017grn,Touboul:2022yrw} have led to stringent constraints on light scalars with couplings $d_X$ as above. In particular, for the linear gluon coupling $d^{(1)}_{g} \lesssim 10^{-3}$ ($10^{-6}$) for all particle masses $m \lesssim 10^{-6}$ ($10^{-14}$) eV (see~\cite{Touboul:2022yrw,Oswald:2021vtc} and refs. therein), for the linear electron coupling $d^{(1)}_{m_e} \lesssim 1$ ($10^{-2}$) for $m \lesssim 10^{-6}$ ($10^{-14}$) eV, and for the linear photon coupling $d^{(1)}_{\a} \lesssim 10^{-1}$ ($10^{-4}$) for $m\lesssim 10^{-6}$ ($10^{-14} \eV$) (see~\cite{Hees:2018fpg} and refs. therein). Constraints on the quadratic couplings are weaker, but as we shall see below, still relevant. 

One can also search for these couplings through direct detection of oscillation of 
fundamental constants 
from the oscillation of the bosonic DM field~\cite{Arvanitaki:2014faa,Banerjee:2018xmn}. 
This variation is characterized at leading order by
\begin{align}
 \frac{\d X}{X_0} \simeq \frac{d_{X}^{(j)}}{j}\frac{\phi^j}{\Mpl^j}
 				\simeq \frac{d_{X}^{(j)}}{j} \left(\frac{\sqrt{2\rho_{\rm DM}}}{m\,\Mpl}\right)^j,
\end{align}
where $\rho_{\rm DM}$ is the density of DM in the vicinity of the experiment, and $j=1$ (2) for linear (quadratic) coupling to $\phi$. 
The typical value for the local density is $\rho_{\rm local} = 0.4$ GeV/cm$^3$, though it can be larger if the field becomes bound to the Earth or Sun \cite{Banerjee:2019epw,Banerjee:2019xuy}.
Substituting Eq. \eqref{eq:dX}, we can write the above equations in a compact form
\begin{align} \label{eq:dXcompact}
 \frac{\d X(\phi)}{X_{0}} &\simeq \frac{N^j \D^{N-j}}{j}\left(\frac{\sqrt{2\rho_{\rm DM}}}{m\,\Mpl}\right)^j \nn \\
 		&\simeq 
		\begin{cases}
		 \displaystyle{2\times10^{-18}\,N \D^{N-1}\left(\frac{10^{-13}\,{\rm eV}}{m}\right)
						\sqrt\frac{\rho_{\rm DM}}{\rho_{\rm local}}} &  \\
						& \hspace{-2cm} ({\rm for} \,\,j=1) \\
		\displaystyle{2\times10^{-36}\,N^2 \D^{N-2}\left(\frac{10^{-13}\,{\rm eV}}{m}\right)^2
						\frac{\rho_{\rm DM}}{\rho_{\rm local}}} &  \\
						& \hspace{-2cm} ({\rm for} \,\,j=2)
		\end{cases}
\end{align}
for $X=m_e,\a,g$, where we have taken $|c_N|\simeq \sin\b\simeq\cos\b\simeq1$. 
For comparison, present experimental sensitivity to $\d m_e/m_{e,0}$ is at the level of $10^{-16}$ for microwave clocks, but somewhat higher for molecular clocks with some prospect to improve to $10^{-21}$ in the coming years; for the $\a$-coupling, current optical clock searches can achieve $10^{-18}$, and a nuclear clock could potentially reach $10^{-23}$ (see \cite{Safronova:2017xyt} and references therein). 
See~\cite{Oswald:2021vtc} for a discussion about the precision probes related to the gluons and quarks couplings.  

For QCD axions, owing to the suppression required to resolve the quality problem, direct searches for quality couplings is challenging. The linear coupling ($j=1$) term in Eq. \eqref{eq:dXcompact} gives
\begin{equation}
 \left(\frac{\d X(\phi)}{X_{0}}\right)_{\rm QCD} \sim 10^{-98} \left(\frac{10^{-13}\,{\rm eV}}{m}\right)
						\sqrt\frac{\rho_{\rm DM}}{\rho_{\rm local}}
\end{equation}
for $f=10^{10}$ GeV ($N=10$), and even smaller for $f=10^{12}$ GeV ($N=13$) and/or for quadratic couplings ($j=2$).
 
The scale of these couplings is exceedingly small, even for ALPs. 
For quadratic couplings ($j=2$), there is a simple expression for the coupling of Eq. \eqref{eq:dX} such that it satisfies the condition $\d m\ll m$ of Eq. \eqref{eq:epALP}:
\begin{equation}
 d_X^{(2)} \ll \frac{m^2}{\Mpl^2} = 10^{-56}\left(\frac{m}{\rm eV}\right)^2,
\end{equation}
which is far out of reach of experimental searches for the foreseeable future. For linear couplings ($j=1$), the condition is more complicated but can be written as
\begin{equation}
 d_X^{(1)} \ll \frac{m^2}{\Mpl^2}\frac{\tan\b\,\D}{N},
\end{equation}
which is suppressed by an additional factor of $\D/N \ll 1$. Therefore, natural couplings are out of reach for now.

\begin{figure}[t]
 \centering
 \includegraphics[scale=0.85]{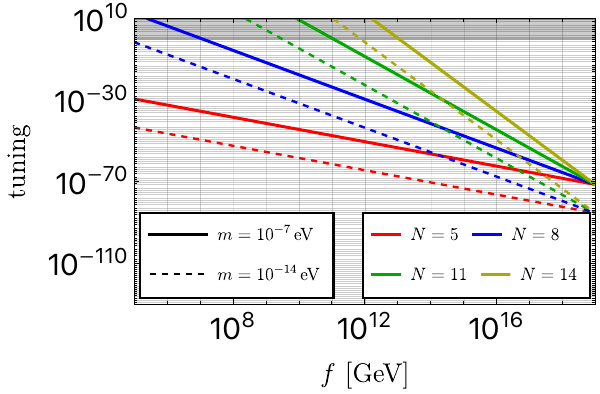}
 \caption{Effective tuning of the ALP theory, given by the RHS of Eq. \eqref{eq:ALPtuning}, for ALP mass $m$ (solid and dashed) and Planck-suppressed operator dimension $N$ (red, blue, green, and yellow) given in the legend, as a function of $f$. The shaded region denotes ``tuning"$>1$.}
 \label{fig:tuning}
\end{figure}

Relative to the case of QCD axions, where additional fine-tuning of the phase parameter $\b'$ spoiled the solution of the strong CP problem (see Section \ref{sec:QCD}), for ALPs the problem is naturalness of the mass. Therefore it is more compelling to ask what level of fine-tuning might be required to produce an ALP with the desired properties. Rather than $\b' \ll 1$, here we may require $\b - \pi/2 \ll 1$ so that $\cos\b \ll 1$. Expanding in this limit, Eq. \eqref{eq:epALP} is equivalent to
\begin{equation} \label{eq:ALPtuning}
  |c_N|\times\left|\b - \frac{\pi}{2}\right| \ll \frac{m^2 f^2}{\Mpl^4} \frac{\Mpl^N}{N^2 f^N},
\end{equation}
i.e. one either tunes $|c_N|\ll 1$ or $|\b - \pi/2|\ll 1$ or both. 
The level of tuning of an ALP theory with a given $N$ and $m$ is given in Fig. \ref{fig:tuning}, where ``tuning'' is defined by the right-hand-side (RHS) of Eq. \eqref{eq:ALPtuning}.
We see that there is a trade-off between the level of tuning in the model (which prefers larger $N$ and smaller $f$) and the possibility of direct detection (which prefers smaller $N$ and larger $f$). 

\begin{figure*}[t]
 \centering
 \includegraphics[scale=0.62]{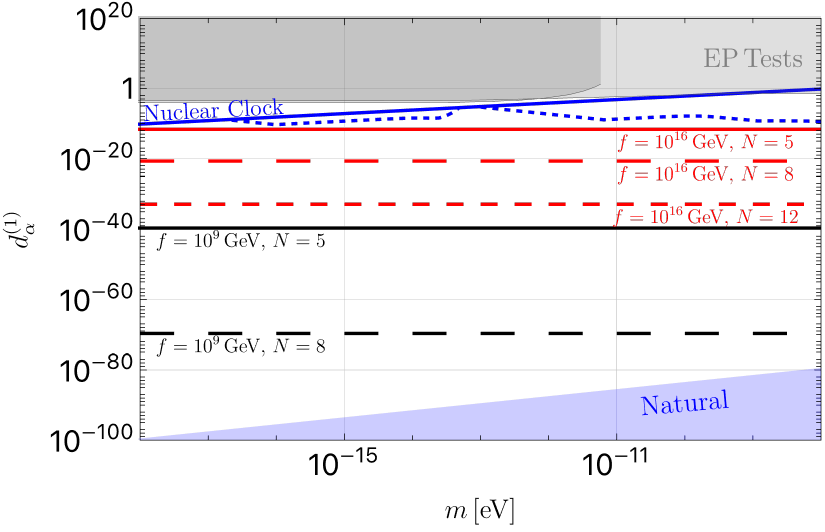}
  \includegraphics[scale=0.62]{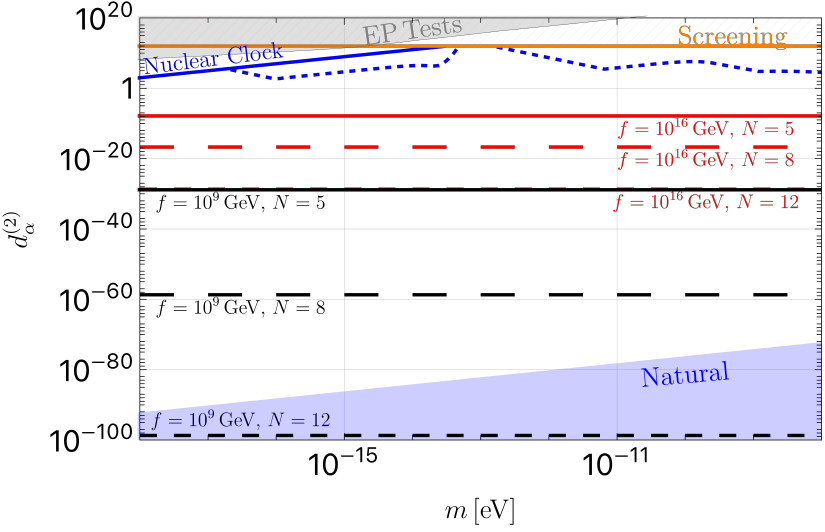}
 \caption{Sensitivity estimate for a nuclear clock with precision at the level of $\d \a/\a \simeq 10^{-23}$, assuming virial DM density (blue thick) and a bound halo around the Earth or the Sun \cite{Banerjee:2019epw,Banerjee:2019xuy} (blue dashed). Also shown are the predicted quality couplings for $5\leq N\leq 12$ (as labelled) and different choices of $f=10^{9}$ GeV (black) and $f=10^{16}$ GeV (red). The gray shaded regions are constrained by EP tests \cite{Smith:1999cr,Schlamminger:2007ht,Berge:2017ovy,Touboul:2017grn,Touboul:2022yrw}, and the blue shaded region is the region of technical naturalness of the scalar field mass.}
 \label{fig:sens}
\end{figure*}

It matters how the tuning of parameters is accomplished. 
If $|c_N| \ll 1$, then all quality couplings $d_X^{(j)}$ are also strongly suppressed (see Eq. \eqref{eq:dX}). If $|\b-\pi/2|\ll1$, then the quadratic couplings become suppressed whereas the linear couplings remain of order $d_X^{(1)} \simeq N\,\D^{N-1}$. Finally, one might imagine a UV model with a bare mass term $m_0^2 < 0$ and a fine cancellation $\d m^2 - |m_0^2| \equiv m^2$ where $m$ is the ultralight mass one searches for in experiment (this is analogous to Higgs fine-tuning); in this case neither linear nor quadratic couplings are necessarily suppressed by the tuning of the theory.

The quality couplings to\footnote{Analogous estimates for other SM operators, e.g. in Eq.~\eqref{eq:Lscalar}, are straightforward. Since neither the couplings Eq.~\eqref{eq:dX} nor the tuning constraint~\eqref{eq:epALP} depend on the SM operator, our estimations of the magnitude of the coupling strength is unchanged in such cases.} 
$\Ocal_{\rm SM} = F^{\m\n}F_{\m\n}$ for $j=1$ (linear coupling to SM) are shown in the left panel of Figure \ref{fig:sens}, and $j=2$ (quadratic couplings to SM) are shown in the right panel. 
The region already ruled out by EP tests is given in grey, and the natural region of coupling space is highlighted in blue. The horizontal lines correspond to Eq.~\eqref{eq:dX} for the labelled values of $N$ and $f$, assuming $c_N \sim \b \sim \Ocal(1)$. We observe that even in the case of a high-density solar halo or Earth halo \cite{Banerjee:2019epw,Banerjee:2019xuy}, a future nuclear clock with precision at the level of $\d\a/\a \sim 10^{-23}$ (blue dashed line) will still not be sufficient to probe these Planck-suppressed couplings.

\section{High-quality QCD relaxion} \label{sec:HQQCD}

We combine elements of $\mathbb Z_N$ QCD axion model with the relaxion, in a way that can ameliorate the challenges described in Section \ref{sec:challenges}. The relaxation of the axion field will preserve the QCD axion solution to the strong CP problem, giving rise to a \emph{low-mass $\mathbb{Z}_N$ QCD axion which also relaxes the electroweak (EW) scale via the relaxion mechanism.}

We again consider $N$ copies of the SM related by a $\mathbb{Z}_N$ symmetry, with an effective potential given in Eq.~\eqref{eq:VN}. We will use the fact that the QCD axion potential depends on the Higgs vev through the quark masses, and thus, it can be used as a trigger for the relaxation of the Higgs mass~\cite{Graham:2015cka, Arkani-Hamed:2016rle}. 
Note that for our purpose, we will only be interested in the shape of the potential and its dependence on Higgs vev. 
 
Starting from a high-energy cut-off $\Lambda$, the EW scale is set by the dynamics of a axion-like field, usually known as a \emph{relaxion}. 
The relaxion-Higgs potential can be written as
\bea
V(\phi,H)&= \left(\Lambda^2-g\Lambda \phi\right)|H|^2+ \lambda |H|^4 \nn \\
		&+ V_{\rm roll}(\phi)+ 
V_{\rm br}(\phi,\left<H\right>)\,, 
\label{eq:relaxion_potential}
\eea
where $V_{\rm roll}=-g\Lambda^3\phi$ ($g$ a dimensionless constant) and $V_{\rm br}$ is called the ``back-reaction'' potential as this back-reacts to the motion of the relaxion and is only active when $\vev{H}\neq 0$. 
In our case, we will take $V_{\rm br} = V_N(\phi)$ in Eq.~\eqref{eq:VN}, which depends linearly on the Higgs vev through $m_u = y_u \vev{H}$, with $y_u$ the Yukawa coupling of the up quark, in contrast to the quadratic case discussed in~\cite{Banerjee:2020kww}. 
Note that with this definition, the SM Higgs vacuum expectation value (vev) would be $\vev{H}=v\sim 174\GeV$. 
See Appendix \ref{app:relaxion} for general details about the relaxion mechanism and constraints. 

In~\cite{Banerjee:2020kww}, the authors discussed the vacuum structure of the relaxion near the EW scale in detail, and showed that the rolling of the relaxion field stops at the first local minimum of the potential it encounters. 
Furthermore, due to the incremental change of the Higgs vev, the relaxion stops at a very shallow part of the potential and thus, its mass is suppressed compared to the naive expected value, a mechanism known as \emph{relaxation} of the axion field.

The relaxion stopping point is determined when the first derivative of $V_{\rm br}(\phi)$ is close to its maximum \cite{Banerjee:2020kww}. 
If, for example, $V_{\rm br}(\phi)\propto \cos(N\theta)$, the relaxion stops around $N \theta_0 \sim 3\pi/2$ for $N$ even, and $N\theta_0 \sim \pi/2$ for $N$ odd. 
In the absence of some breaking of the $\mathbb{Z}_N$ symmetry, this leads to $\mathcal{O}(1)$ CP violating phase in all sectors, and is thus ruled out by the neutron EDM experiments unless $N\sim \mathcal{O}(10^{10})$. 

In order to successfully solve the strong CP problem, we need to find at least one sector in which effective $\theta_0\lesssim \mathcal{O}(10^{-10})$ and identify this sector as our SM (this amounts to a linear tuning of $1/N$).
To do that, we will break the $\mathbb{Z}_N$ symmetry explicitly in the $k=0$ sector by a small parameter, \footnote{We denote all quantities in the $k=0$ sector with a $()^\prime$.} by requiring $y'_u>y_u^{\rm SM}\sim 10^{-5}$; as a result, the confinement scale $\L_{\rm QCD}'$ would change as well~\cite{Berezhiani:2000gh,Hook:2014cda,Fukuda:2015ana}. A possible second source of $\mathbb{Z}_N$ breaking is a change to the Higgs vev in the $k=0$ sector, where we assume $v' \geq v$. We parameterize these two sources of breaking using the parameters
\begin{align} \label{eq:breaking}
 \epsilon_b \equiv  \frac{y_u \Lambda_{\rm QCD}^3}{y'_u \Lambda_{\rm QCD}'^3}\,,
 \qquad \gamma \equiv \frac{v}{v'}\,.
\end{align}
Note that $0 \leq \{\e_b,\g\} \leq 1$, and the $\mathbb{Z}_N$ symmetry is restored for $\e_b,\g\to 1$.

\subsection{Toy Model}
  
We begin with a simplified example where the backreacion potential in the prime sector is of the form $V_{\rm br}(\phi) \propto \cos(\phi/f)$, to illustrate how the mechanism works. In a realistic scenario, one must account for corrections to the backreaction potential in the ${\rm QCD}'$ sector, which we consider in Section \ref{sec:NLO}.

In the toy model, we take the back-reaction potential to be written as
\bea
V_{\rm br}(\phi) \sim &&-\Lambda_{\rm QCD}'^3  y'_u v'  
                \cos\left(\frac{\phi}{f}\right)\left(1 - \e_b \g\right) \nn \\ 
		&& - \Lambda_{\rm QCD}^3 \,y_u v\, \k \,
		        \cos\left(\frac{N\phi}{f}\right)\,,
\label{eq:QCD_backreaction}
\eea
where we ignore $\bar{\th}$ for the purposes of this section. 
If $\epsilon_b \g\ll 1$, we can treat the term proportional to $\cos(N\phi/f)$ as a perturbation and the relaxion stopping point can be written as $\theta_0=\pi/2-\delta_\theta$, 
with 
\bea
\delta_{\theta} \simeq \frac{\epsilon^2}{8}+ N^2 \kappa \epsilon_b\,\gamma \pm \delta - \Ocal(\delta^2)\,,
\label{eq:dtheta_QCD_relaxion}
\eea
where we define 
\bea 
\epsilon^2 &\equiv& \frac{\Lambda_{\rm QCD}'^3  y'_u}{v'^3} = \frac{\L_a^4}{v^4}\frac{\g^3}{\e_b}\,, 
 \\ 
\delta &\equiv& \frac{\e\,v'}{\L} = \frac{\L_a^2}{v^2}\,\left(\frac{v}{\L}\right) \sqrt{\frac{\g}{\e_b}}\,. 
\eea
We are assuming that 
the Higgs mass is relaxed starting from some cut-off $\Lambda$ to the value $v'\gtrsim v$ in the $k=0$ sector, and to $v$ in the SM sector. This amounts to a fine-tuning of order $\gamma^2 = (v/v')^2$.

The relaxion stopping point would be close to $\pi/2$ in the $k=0$ sector which dominates the relaxion potential. 
However in the $k$th sector the stopping point would be shifted by $2\pi k/N$ as per the structure of the potential as seen in Eq.~\eqref{eq:VtotRR}. 
So, if we identify the SM at the $k_{\rm SM}=(3N/4)$-th sector (which is shifted from the dominating sector by 
$2\pi k_{\rm SM}/N = 3\pi/2$), then in our SM the effective theta angle would be $\theta_0 + 3 \pi/2\sim \delta_\theta$. 
We reiterate that selecting the SM out of $N$ sectors as the one with minimum at $3\pi/2$ amounts to tuning of $1/N$. This also implies the constraint that $N$ be a multiple of $4$ in this model, i.e. $N\,{\rm mod}\,4 = 0$. In this sense, the underlying symmetry of this theory is $\mathbb{Z}_{4\Ncal}$, with $\Ncal \equiv N/4$.\footnote{See also \cite{Chen:2021haa} for a GUT-related motivation for this symmetry.}

In order to solve the strong CP problem successfully one requires $\d_\th \lesssim \th_{\rm CP} = \theta_{10}10^{-10}$, the limit on the CP violating phase from neutron EDM experiments; at present, $\th_{10} = 1.4$~\cite{Abel:2020pzs} but is expected to improve in the future \cite{Filippone:2018vxf}. 
To avoid any additional tuning, one would expect each term in $\delta_\theta$, defined in Eq.~\eqref{eq:dtheta_QCD_relaxion}, are separately less than $\th_{\rm CP}$. 
This implies additional conditions, namely
\bea \label{eq:epsqu}
	\e^2 \lesssim \th_{\rm CP} &\Rightarrow& \frac{\g^3}{\e_b} \lesssim 1800\,\th_{10}\,, \\
	\d \lesssim  \th_{\rm CP} &\Rightarrow& \frac{\g}{\e_b} \lesssim 
					6 \left(\frac{\L}{10^6\,{\rm GeV}}\right)^2\th_{10}\,, \label{eq:delta} \\		
	\e_b\g \lesssim \frac{\th_{\rm CP}}{N^2 \k} &=& \frac{ \theta_{10}10^{-10}}{N^2 z^{N-1}} \sqrt{\frac{\pi N}{1-z^2}}\,. \label{eq:kappaconstraint}
\eea
where we have taken $\L_a/v \simeq 5\times10^{-4}$.

Note that, in additional to the QCD and relaxion parameters, we have two additional free parameters $\epsilon_b$ and $\g$, constrained by three inequalities. 
The cut-off of the Higgs mass $\Lambda$ is also constrained by the consistency of the effective theory as $f \gtrsim \Lambda\gtrsim 4\pi v'$. 
Other constraints from the success of the relaxion mechanism are described in Appendix \ref{app:relaxion}; the upshot is that a successful relaxation of the EW scale requires the additional condition 
\begin{equation} \label{eq:conrel}
 \frac{\e_b^{5/2}}{\sqrt{\g}} \lesssim 24\pi^2 \left(\frac{\L_a^{10} \Mpl^4}{\L^{11} v^3}\right).
\end{equation}
Finally, as mentioned above, we must ensure that the values of $\e_b$ and $\g$ are consistent with the change of the QCD scale as the Higgs vev changes, i.e. $\L_{\rm QCD}' > \L_{\rm QCD}'(\gamma)$, which roughly translates to the constraint \cite{Berezhiani:2000gh,Hook:2014cda,Fukuda:2015ana}
\begin{equation} \label{eq:conYana}
\e_b \lesssim \gamma\,.
\end{equation} 

\subsection{Model including NLO corrections}
\label{sec:NLO}

In the toy model above, we only consider the first term of the leading order (LO) ${\rm QCD}'$ potential.  
However, there are additional terms in the potential at LO, as well as the higher-harmonic contributions coming from the non-leading-order (NLO) terms. Both of these contributions can shift the vacuum of the SM sector and spoil the mechanism.\footnote{We are grateful to Javi Serra, Stefan Stelzl, and Andreas Weiler for clarifying this point.} In this section we extend our analysis to include these additional terms.

First, note that as the relaxion stopping point is determined when the
first derivative of $V_{\rm br}(\phi)$ is close to its maximum, the full LO ${\rm QCD}'$ potential would lead to the relaxion stopping point
of $\theta_0 \simeq \cos^{-1}[z']$, where $z'=m_u'/m_d'$. 
This suggests that a stopping point close to $\pi/2$ would lead to $\mathcal{O}(1)$ $\theta_{\rm CP}$. 

Furthermore, to obtain a $\mathbb Z_{N}$-symmetric potential at LO, one needs to subtract a term of the form of $\Lambda_{\rm QCD}^3 \,y_u v\, 
\sqrt{1+z^2+2 z\cos(\theta_a)}$ (leading order expansion of this term is shown by the second term of the first line in Eq.~\eqref{eq:QCD_backreaction}). 
For $z'\neq z$, this term also shifts $\theta_0$ by an amount proportional to the ratio of the amplitudes of QCD and ${\rm QCD}'$ potentials and the phase misalignment as shown in Eq.~\eqref{eq:appNLOzz} (see Appendix~\ref{app:QCD_prime} for a detailed discussion).   

NLO contributions to the potential are not aligned to the LO contribution, and induce a shift to the stopping point of $\mathcal{O}(R\, m_{u,d}'/\Lambda_{\rm QCD}')$ where $R\sim \mathcal{O}(10^{-3})$ is the NLO coefficient~\cite{Lu:2020rhp}. 
Thus, to successfully solve the strong CP problem we must require that $\mathcal{O}(R\, m_{u,d}'/\Lambda_{\rm QCD}')\lesssim \theta_{\rm CP}$, which in turn requires a large hierarchy between the SM and the hidden sector confinement scales, as shown in 
Eqs.~(\ref{eq:appNLOmq}-\ref{eq:LQCDcon}). 
One mechanism to accomplish this is to introduce additional heavy vector-like fermions, as discussed in Appendix~\ref{app:large_alphas}.

Below we discuss two variants of the toy model, which lead to suppression of the above discussed higher-order contributions. These require additional breaking of the $\mathbb{Z}_N$, in addition to some flavor symmetry of the hidden sector. We analyse the phenomenology of the above variants, in detail, in Appendix \ref{app:QCD_prime}. Below, we summarize the main results and how these terms affect the parameter space of the model.

 In the first, we consider the case where the prime ($k=0$) sector possesses effective isospin or $\mathbb{Z}_2^{\rm flav}$, such that $z'\equiv y_u'/y_d'=1$, which is sometimes denoted as natural flavor conservation.\footnote{For a discussion related to such constructions see {\it e.g.}~\cite{Buchmuller:1985jz,Hall:1993ca, Branco:2011iw} and Refs. therein.} In this case, the potential of the prime sector is minimized at $\phi/f \simeq \pi - \d_\th$; as in the toy model, we require each term in $\d_\th$ be smaller than the CP angle $\theta_{\rm CP}$. 
The $k$th sector will be shifted from the prime sector by $2\pi k/N$, so in order to identify the SM with a sector having $\theta_0 \simeq 0$, one needs an underlying symmetry of $\mathbb{Z}_{2\Ncal}$ with $\Ncal = N/2$. 
In this case, in addition to Eqs.~(\ref{eq:epsqu}-\ref{eq:conYana}), there are two new constraints (see Eqs.~(\ref{eq:appzz}-\ref{eq:appmqlam}) in Appendix \ref{app:QCD_prime}):
\begin{align}
    \e_b\g &\lesssim \th_{\rm CP}\frac{1-z}{z}, \label{eq:appNLOzz}\\
    \frac{R}{\e_b\g} &\lesssim 
            \left(\frac{\L_{\rm QCD}'}{\L_a}\right)^4 \th_{\rm CP} \label{eq:appNLOmq},
\end{align}         
where $R\sim \Ocal(10^{-3})$ is an NLO suppression factor \cite{Lu:2020rhp}. 
Note that the constraint \eqref{eq:appNLOzz} is strictly stronger than \eqref{eq:kappaconstraint} for any $N\gtrsim 2$.
For $\L_{\rm QCD}' \lesssim 1$ TeV, the above constraints can be simultaneously satisfied, for  $\th_{\rm CP}\gtrsim 2\times 10^{-10}$ and $z\simeq 0.48$ as usual (see Eq.~\eqref{eq:LQCDcon}).

In the second variant, sketched below, we assume that the hidden sector has an extended flavor symmetry leading to $z' = 1/2$. In this case the minimum of the effective potential is shifted to $2\pi/3$. This leads to a further suppression of the SM corrections to the stopping point, as detailed in Appendix \ref{app:QCD_prime}. 
The upshot is that one can obtain the results for this case using the substitution $z/(1-z) \to 2C_z \equiv (z/2)(2-5z+2z^2)/(1-z+z^2)^{3/2}$ above; for the central value $z=0.485$~\cite{Fodor:2016bgu}, one obtains $C_z \simeq 8\times10^{-3}$, whereas for the $2\s$  limit (up) $z = 0.504$ one obtains $C_z \simeq 2 \times10^{-3}$ and for (down) $z = 0.466$ one obtains $C_z \simeq 1.9\times10^{-2}$. In this case, the underlying symmetry must be $\mathbb{Z}_{3\Ncal}$ with $\Ncal = N/3$. Given these inputs, the constraint \eqref{eq:appNLOzz} is weaker than \eqref{eq:kappaconstraint} unless $N\gtrsim 9,\,12,\,15$ for $z=0.466,\,0.485,\,0.504$. 

We now briefly sketch how to construct a flavor model, leading to $z' = 1/2$. This is based on the model discussed in~\cite{Balkin:2021rvh}, which generalizes the $\mathbb{Z}_2^{\rm flav}$ symmetry to $\mathbb{Z}_5^{\rm flav}\,,$ that can be realized in extra dimensional constructions. 
The idea is that there is a 5-plet $\psi_i$ ($i=1..5$), with $\psi_1=u$, identifies with the up quark hidden singlet field, and the rest of the four components of $\psi_i$ carry down-singlet hidden quark quantum numbers. 
Assuming that the rest of the hidden sector fields are singlets, and the only light down field is made of an equal linear combination of the four $\psi_i$s, one would obtain a model in which the effective hidden down quark Yukawa for the light field is $\sqrt{4}$ times the up one, as required above (for more detail see~\cite{Balkin:2021rvh}).

We can combine the new constraints Eq.~\eqref{eq:appNLOzz} and \eqref{eq:appNLOmq} in order to set a lower bound on $\L_{\rm QCD}'$ in the model:
\begin{align} \label{eq:LQCDcon}
 \Lambda_{\rm QCD}' \gtrsim \L_a \left(\frac{(r_1+r_2)z}{\th_{\rm CP}^2(1-z)}\right)^{1/4}.
\end{align}
In what follows, we always choose $\L_{\rm QCD}'$ large enough that it satisfies this limit. 

As discussed in~\cite{Berezhiani:2000gh,Fukuda:2015ana} and Appendix \ref{app:large_alphas}, we can only obtain $\L_{\rm QCD}'\sim \mathcal{O}(\GeV)$, for $\gamma=v/v'\sim 10^{-3}$. Thus, to achieve a larger $\L_{\rm QCD}'$, we need to introduce new states both in the SM and the hidden sectors which are charged under $SU(3)_C$. 
The mass difference of these states explicitly breaks the $\mathbb{Z}_{N}$ symmetry. Thus, in order for the $\mathbb{Z}_{N}$ to be a good symmetry for the axion, we require that the Peccei-Quinn symmetry breaking scale, $f$, is higher than the mass of the heaviest new state. 
For our construction this can be achieved for $f\gtrsim \mathcal{O}\left(10^{12} \GeV\right)$. 

Intriguingly, the parameter space consistent with all other constraints \eqref{eq:delta}, \eqref{eq:conrel},\eqref{eq:appNLOzz}, and $\L \gtrsim 4\pi v'$, is exceedingly predictive. 
The largest $v'$ allowed for a given $\th_{\rm CP}$ can be determined by the intersection of ~\eqref{eq:delta}, \eqref{eq:appNLOzz}, and saturating $v'=\L/4\pi$, which gives
\begin{equation} \label{eq:vpint}
    v' \lesssim v\left[\frac{1}{(4\pi)^2\th_{\rm CP}^3}\left(\frac{\L_a}{v}\right)^4\frac{z}{1-z}\right]^{1/4}.
\end{equation}

Note that the intersection of Eqs.~\eqref{eq:delta}, \eqref{eq:conrel}, and \eqref{eq:appNLOzz} implies
\begin{equation}
    v' \gtrsim \left[\left(\frac{z}{1-z}\right)^3\frac{1}{\theta_{\rm CP}^{14}}\frac{\L_a^{12} }{24\pi^2\,\Mpl^4} \right]^{1/8}.
\end{equation}
In the $z'=1$ model, $\th_{10} = 1$ (10), the vev in the prime sector can be as small as $v'_{\rm min} = 2400$ ($45$) TeV, whereas in the $z'=1/2$ model using the $2\s$ value of $z$, we find $v'_{\rm min} = 314$ ($5.6$) TeV. 

Another way to explore the parameter space is to inquire into the lowest allowed value of $\th_{\rm CP}$.
The two above constraints on $v'$ are simultaneously saturated at the minimum allowed $\th_{\rm CP}$, which is
\begin{align} \label{eq:thcpmin}
    \th_{\rm CP,min} &= \left[\frac{2^5\pi^2}{3}\frac{z}{1-z}\right]^{1/8}\sqrt{\frac{\L_a}{\Mpl}} \nn \\
        &\simeq 3.33\times10^{-10}
                \left(\frac{z}{1-z}\right)^{1/8}.
\end{align}
Substituting this result into Eq.~\eqref{eq:vpint}, the corresponding Higgs vev in the prime sector is
\begin{align} \label{eq:vpCPmin}
    \frac{v'}{v}\Big|_{\th_{\rm CP,min}} &= \frac{\L_a^{5/8}\Mpl^{3/8}}{2\,v}\left(\frac{3^3}{2^{15}\pi^{22}}\right)^{1/32}\left(\frac{z}{1-z}\right)^{5/32} \nn \\
        &\simeq 1750\left(\frac{z}{1-z}\right)^{5/32}.
\end{align}
For the $z'=1/2$ model, as before, $z/(1-z) \to 2C_z \approx 4\times 10^{-3}$ (for the up $2\s$ value of $z$), implying $\th_{\rm CP,min} \simeq 1.67\times10^{-10}$.

The strongest limit on $\theta_{\rm CP}$ comes from the non-observation of a neutron electric dipole moment; the current upper bound is $\th_{\rm CP} \lesssim 1.4\times 10^{-10}$ ($90\%$ confidence) \cite{Abel:2020pzs}. We have shown above that our model can achieve $\theta_{\rm CP}/10^{-10} \sim 1-2$ at the lowest, implying some tension with existing constraints. 
Note that $C_z\to 0$ for $z\to 1/2$, and in this case $\th_{\rm CP}$ is strongly suppressed (see Eq.~\eqref{eq:thcpmin}); this value $z=1/2$ lies in the $2\sigma$ window of the current best-fit value of $z$ \cite{Fodor:2016bgu}. 
The PDG value of $z$ has a larger error bar compared to those used here~\cite{Workman:2022ynf}, and leads to the same conclusion.

 \begin{figure*}
\centering
\begin{minipage}[c]{0.45\textwidth}
\hspace{-7cm} \large{(a)} \\
\vspace{-0.5cm}
\includegraphics[width=\linewidth]{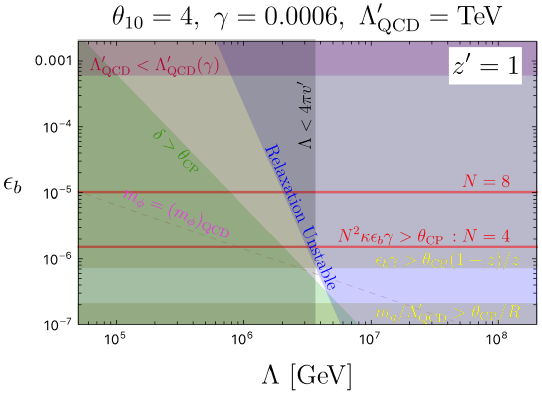}
\end{minipage}
\hfill
\begin{minipage}[c]{0.45\textwidth}
\hspace{-7cm} \large{(b)} \\
\vspace{-0.5cm}
\includegraphics[width=\linewidth]{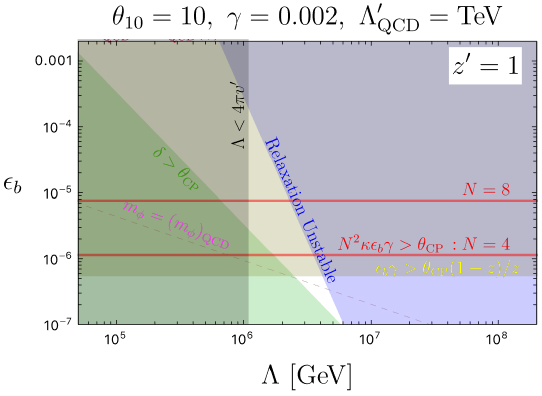}
\end{minipage}
\\
\vspace{0.2cm}
\begin{minipage}[c]{0.45\textwidth}
\hspace{-7cm} \large{(c)} \\
\vspace{-0.5cm}
\includegraphics[width=\linewidth]{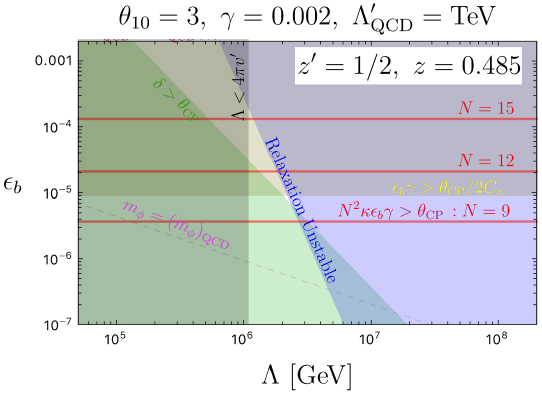}
\end{minipage}
\hfill
\begin{minipage}[c]{.45\textwidth}
\hspace{-7cm} \large{(d)} \\
\vspace{-0.5cm}
\includegraphics[width=\linewidth]{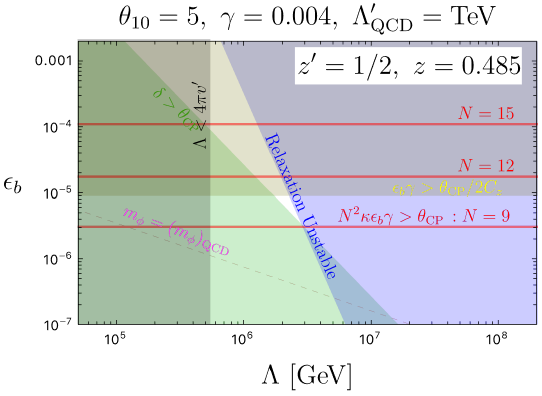}
\end{minipage}
\\
\vspace{0.2cm}
\begin{minipage}[c]{0.45\textwidth}
\hspace{-7cm} \large{(e)} \\
\vspace{-0.5cm}
\includegraphics[width=\linewidth]{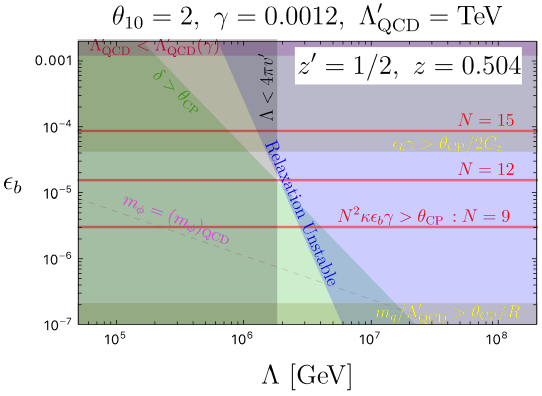}
\end{minipage}
\hfill
\begin{minipage}[c]{.45\textwidth}
\hspace{-7cm} \large{(f)} \\
\vspace{-0.5cm}
\includegraphics[width=\linewidth]{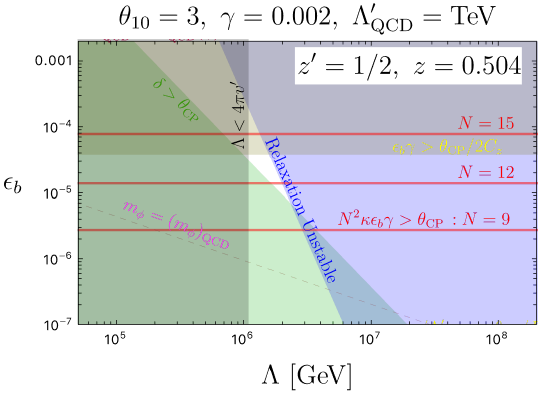}
\end{minipage}
\caption{Parameter space for the $\mathbb{Z}_N$ QCD relaxion, using both the $z'=1$ model (panels a and b) and the $z'=1/2$ model (c - f). For the purpose of illustration, we fix $\L_{\rm QCD}' = $TeV, though smaller values are possible (see Eq.~\eqref{eq:LQCDcon}); we vary both $\th_{10}$ and $\g$ as labelled in each panel. 
The two symmetry-breaking parameters $\e_b$ and $\g$
are defined in Eq.~\eqref{eq:breaking}.
The constraints shown in green, red, blue, and purple are given in Eqs.~(\ref{eq:delta}-\ref{eq:conYana}) (respectively); the higher-harmonic constraints in Eqs.~(\ref{eq:appNLOzz}-\ref{eq:appNLOmq}) are shown in yellow, and $\L<4\pi v'$ is given in black. 
 The dashed line illustrates where the relaxion mass is equal to $(m_\phi)_{\rm QCD} \equiv \L_a^4/f^2$; above the line, $m_\phi < (m_\phi)_{\rm QCD}$. }
\label{fig:HQQCD}
\end{figure*}

In Figure \ref{fig:HQQCD}, we illustrate the parameter space for each of the above models: $z'=1$ (panels a and b); and 
$z'=1/2$ with $z=0.485$ (c and d) or with $z=0.504$ (e and f). 
(The individual constraints are described in the figure caption.)
As noted above, the parameter space is exceedingly predictive, as we find the cut-off scale $\L \simeq 10^6-10^7$ GeV and $\th_{10} \gtrsim 1.6$. 
Note that the $\e^2$ inequality \eqref{eq:epsqu} does not appear, as this constraint is always much weaker than the others. 

After the rolling stops, the mass of relaxion will be relaxed to a value modified from the naive expectation; it can be written as
\bea
m_\phi^2 &=& \frac{\Lambda_{\rm QCD}'^3 y'_u v'}{f_a^2}\times \delta \, \nn \\
		&=& (m_\phi^2)_{\rm QCD} \frac{\d}{\e_b\g}\,,
\eea
where $(m_\phi^2)_{\rm QCD} \equiv \L_a^4/f^2$. 
In particular, it is suppressed by $\d \ll 1$ but enhanced by $(\e_b\g)^{-1} \gg 1$ relative to the naive QCD expectation. 
Note that one can express $\delta$ in terms of theory parameters as
\bea
\!\!\!\!\!\!\!\!\!\!\!\!\!\!\!\!\!\!\!\!\!\delta 
&\simeq& 4\times 10^{-11}\left(\frac{10^6\,{\rm GeV}}{\Lambda}\right)\sqrt{\frac{\g}{\epsilon_b}}\,.
\eea
So, finally the (rel)-axion mass can be written as,
\bea
\!\!\!\!\!\!\!\!\!\!\!\!\!\!\!\!\!
\frac{m_\phi^2}{(m_\phi^2)_{\rm QCD}} \simeq 1.3\times 10^{-7} \left(\frac{10^{-7}}{\e_b^3 \g}\right)^{1/2}
				\left(\frac{10^6\,{\rm GeV}}{\Lambda}\right)\!.  
\eea
In Figure \ref{fig:HQQCD}, the pink dashed lines denote the parameter space satisfying $m_\phi \simeq (m_\phi)_{\rm QCD}$.

\subsection{Quality of the $\mathbb{Z}_{N}$ QCD relaxion}

As before, we combine the low-energy axion potential with that induced by Planck-suppressed operators in Eq.~\eqref{eq:PQB2} to see whether the latter will spoil the solution to the strong CP problem. The combined potential at leading order is
\begin{align}
 V(\phi) &= |c_N|\D^N \Mpl^4 \cos\left(\frac{N\,\phi}{f} + \d'\right) 
 			- m_\phi^2 f^2\,\cos\left(\frac{\phi}{f}\right) \nn \\
	&= |c_N|\D^N \Mpl^4 \cos\left(\frac{N\,\phi}{f} + \d'\right) 
			- \frac{\d\,\L_a^4}{\e_b\g}\,\cos\left(\frac{\phi}{f}\right).
\end{align}
The first derivative is
\begin{align}
 0 &= V'(\vev{\phi}) 
 				\approx |c_N|\D^N N\,\Mpl^4 \sin\d' + \frac{\d\,\L_a^4}{\e_b\g} \e\,,
\end{align}
which implies the constraint
\begin{equation} \label{eq:epRelaxion}
 |\e| = \left|\frac{|c_N| \sin\d'\,\D^N \,N\Mpl^4}{\L_{a}^4}\frac{\e_b\g}{\d}\right| \lesssim 10^{-10}.
\end{equation}

The constraint in Eq.~\ref{eq:epRelaxion} is identical to Eq.~\eqref{eq:epZN} except for the additional factor of $\e_b\g/\d$ on the left-hand-side. 
This factor is at most $10^3$ in the parameter space we consider, and the constraint on $N$ depends on it logarithmically. Therefore, for our purposes we can treat the quality of our QCD relaxion as very similar to the $\mathbb{Z}_N$ axion considered in Refs. \cite{Hook:2018jle,DiLuzio:2021pxd,DiLuzio:2021gos} (see black dashed line in Fig. \ref{fig:fmax}).

As discussed above, one way to realize our mechanism is to introduce new heavy fermions which are charged under $SU(3)_C$. These heavy states contribute to additional $\mathbb{Z}_N$-breaking, leading to the additional requirement that the decay constant of the QCD relaxion be heavier than such states, e.g. $f\gtrsim 10^{12}$ GeV. Comparing to Fig. \ref{fig:fmax}, this would imply a lower limit on $N$ in our model of order $N_{\rm min} = 12$. This result depends on the mechanism for enhancing $\L_{\rm QCD}'$, and thus there may be ways to modify the model to achieve a lower $f$, and therefore a lower $N$, consistent with the mechanism.

\subsection{Direct searches for the $\mathbb{Z}_{N}$ QCD relaxion}
\label{sec:relaxedpheno}

Here we outline the phenomenological implication of our QCD relaxion. 
In our model, the axion has a CP-violating phase of $\delta_\theta$. 
Like the usual relaxion models, due to the relaxion-Higgs mixing angle 
\bea
\sin\theta_{h\phi} \simeq \frac{\L_a^4}{v^3 f}\delta_\theta\,,
\eea
the QCD-axion also has scalar interaction with the SM. See~\cite{Flacke:2016szy,Frugiuele:2018coc,Banerjee:2020kww} for a detailed discussion of relaxion phenomenology. 

The QCD axion also induces a linear scalar interaction with nucleons $(N)$, through the pion-nucleon sigma term in the presence of a CP-violating phase as~\cite{Moody:1984ba,Gasser:1987rb,Bernard:1992qa,Scherer:2012xha}  
\bea
\mathcal{L}\supset - g_{\phi NN}\phi \bar{N}N\,,
\eea
where $g_{\phi NN}$ denotes the scalar coupling strength of the QCD axion with the nucleons. 
Using $\partial \ln m_N/\partial \ln m_\pi^2\simeq 0.06$~\cite{ Hoferichter:2015tha,Kim:2022ype}, and $m_um_d/(m_u+m_d)^2\simeq 0.22\,$,
we obtain 
\bea
g_{\phi NN}\simeq 1.3 \times 10^{-2}\,\frac{m_N}{f}\delta_\theta\,, 
\eea
where $\delta_{\theta}\sim \mathcal{O}(10^{-10})$ is the total CP violating phase in our case as discussed in Eq.~\eqref{eq:dtheta_QCD_relaxion}. 
In our model, the range of the total CP violating phase is $1\lesssim (\delta_\theta/10^{-10})\lesssim 10$, thus the strength of the scalar interaction of the QCD-axion to the SM is
\bea
\!\!\!\!\!\!\!\!\!\!
\frac{ 10^{-24}}{f_{12}}\lesssim g_{\phi NN}
            \lesssim \frac{10^{-23}}{f_{12}}\,,
        \label{eq:bound_scalar_coupling}    
\eea 
where, $f_{12}=f/(10^{12}\GeV)$ and we have used $m_N\sim 1\GeV$. The strongest bound on $g_{\phi NN}$ comes from the experiments looking for the existence of fifth force and/or violation of equivalence principle (EP)~\cite{Smith:1999cr,Schlamminger:2007ht,Berge:2017ovy,Touboul:2017grn,Touboul:2022yrw}. The bound from EP violation searches, for the axion mass around $10^{-6}\eV$, is $g_{\phi NN}\lesssim 10^{-21}$, which becomes stronger as we go to the lower masses. 
Note that, in our model, the mass of the QCD relaxion is slightly lighter than the QCD axion. Thus, for a given $f$, one should be careful about analysing the EP bounds.  

The QCD axion also has pseudoscalar interaction with the SM fermions as, $\mathcal{L}\supset -g_{\psi}^{a} a \bar\psi i \gamma^5\psi$ with $g_{\psi}^a = C_\psi m_\psi/f$. The coefficient $C_\psi$ depends on QCD axion models~\cite{Weinberg:1977ma,Wilczek:1977pj,Dine:1981rt,Zhitnitsky:1980tq,Kim:1979if,Shifman:1979if,GrillidiCortona:2015jxo}. 
Many experimental efforts are concentrated on probing QCD axion through its pseudoscalar interaction with the SM (see e.g.~\cite{Graham:2013gfa} and Refs. therein). 
In our model, the product of the scalar and the pseudoscalar coupling of the QCD relaxion to the nucleon can be written as 
\bea
g_{N}^{a}g_{\phi NN} = 1.3 \times 10^{-2}\,\frac{C_N\, m_N^2}{f^2}  \delta_\theta \,,
\eea 
where $C_N$ is some model dependent coefficient of the nucleons arising from the pseudoscalar interaction of the axion to protons and/or neutrons~\cite{GrillidiCortona:2015jxo}. 
In our model, the strength of the axion-nucleon scalar interaction is bounded and using Eq.~\eqref{eq:bound_scalar_coupling} one can more specifically limit the product of axion-proton pseudoscalar and axion-nucleon scalar coupling as, 
\bea
\frac{6\times 10^{-37}}{f_{12}^2} \lesssim |g_{p}^{a} g_{\phi NN}| 
                \lesssim \frac{6 \times 10^{-36}}{f_{12}^2}\,.
\eea 
Note that, in the above estimate we use the axial coupling strength of proton $C_p=-0.47$ which is obtained in the KSVZ QCD axion model. Another QCD axion model such as DFSZ may provide a different value of $C_p$~\cite{GrillidiCortona:2015jxo}. 
The above parameter range will be probed by the ARIADNE experiment whose projected reach is 
$|g_{p}^{a} g_{\phi NN}|\lesssim 10^{-38}- 10^{-37} $ for $f\sim 10^{12}\GeV$~\cite{Arvanitaki:2014dfa,ARIADNE:2017tdd}. 

The QUAX experiment is also looking for similar scalar-pseudoscalar interaction, using the pseudoscalar electron coupling $g_e^a$ rather than $g_p^a$. 
They provide the current constraint on $|g_{e}^{a} g_{\phi NN}|\lesssim 5.7\times 10^{-32}$ in the mass range of $10^{-5}\gtrsim m_\phi/\eV\gtrsim 6\times 10^{-13}$ by updating their previous result by $\mathcal{O}(10^{2})$~\cite{Crescini:2017uxs,Crescini:2020ykp}. We estimate the range of $|g_{e}^{a} g_{\phi NN}|$ in our model as 
\bea
\frac{2\times 10^{-40}}{f_{12}^2}\lesssim |g_{e}^{a}g_{\phi NN}| 
            \lesssim \frac{2 \times 10^{-39}}{f_{12}^2}\,, 
\eea
where we use $C_e=1/3$; this parameter is model-dependent, and this value is on the larger side of model-parameter possibilities \cite{DiLuzio:2020wdo}. Although our predicted range is beyond the current experimental reach, our model presents an opportunity for scalar and pseudoscalar searches to work together to confirm (or refute) the existence of such axions in a complementary way.

\section{Discussion}

In this work we analysed how Planck-suppressed (quality) operators affect the low-energy dynamics of theories involving QCD axions or axion like particles (ALPs). 
For the QCD axion, the quality operators lead to the well-known \textit{QCD axion quality problem}, whereas for ALPs, they may lead to an equally severe fine-tuning problem. 
Quality operators also induce scalar interaction between the Standard Model (SM) fields and QCD axions/ALPs. 
In the absence of CP violation, we obtain SM-ALP scalar interaction in quadratic order of the ALP field, whereas if CP is broken by gravity, ALP-SM scalar interactions are generated even at linear order. 
These interactions can be probed by various precision experiments.
The strength of the scalar and pseudoscalar interactions are closely related, and therefore these search strategies can complement one another.

We also provide a framework for addressing both the Higgs hierarchy and the strong CP problems together. We invoke a relaxation mechanism where the Higgs mass is scanned during inflation and the QCD axion plays the role of the relaxion. 
We show that a $\mathbb Z_{N}$-symmetric back-reaction potential which is broken explicitly by a small parameter can address both of these problems simultaneously. 
Depending on the symmetry of the dominant sector, one can accomplish this mechanism with underlying symmetries of $\mathbb{Z}_{2\Ncal}$ or $\mathbb{Z}_{3\Ncal}$, with interesting implications for the resulting parameter space.
We show that one of the sectors, identified with the SM, has effective CP violating phase $\theta_0\lesssim \mathcal{O}(10^{-10})$. The tuning in the model is linear and of $\mathcal{O}(\mathcal N)$. Our model cannot fully ameliorate the hierarchy problem, as it leaves a little hierarchy to address. 
The mass of the QCD relaxion obtained in our model can also be lighter than that of the canonical QCD axion.

Our model can accommodate a CP-violating phase of $1 \lesssim \th_0/10^{-10} \lesssim 10$. This range of CP-violating phase is already being probed by neutron electric dipole moment experiments \cite{Abel:2020pzs}, and will be fully probed within the next five years~\cite{Filippone:2018vxf}. 
Due to the underlying $\mathbb Z_{N}$ symmetry which can be gauged, this model exhibits better protection against quality operators than the vanilla QCD axion/relaxion models. 
Due to the predicted narrow range of CP-violating phase, our model can also be used as target of experiments like ARIADNE \cite{Arvanitaki:2014dfa,ARIADNE:2017tdd} and/or QUAX \cite{Crescini:2017uxs,Crescini:2020ykp}, which are sensitive to the product of scalar and pseudoscalar interaction of the QCD axion to the SM. 
In the case of QCD relaxion dark matter, precision searches can also be applied. 
However, further investigation of this possibility is beyond the scope of the current work.

\section*{Note added}

As this work was finalized, \cite{Chatrchyan:2022pcb} appeared, which also address issues related to the QCD relaxion by changing the relaxion evolution during inflation. We also note the recent work \cite{Zhang:2022ykd}, which discusses the relationship of Planck-suppressed operators and fifth forces.

\section*{Acknowledgements}

We thank Kfir Blum for useful discussions, as well as Javi Serra, Stelzl Stefan, and Andreas Weiler for critical comments on an early version of this work. AB and GP acknowledge MITP and the Aspen Center for Physics for their hospitality and support where the part of the work was performed.  
The work of AB is supported by the Azrieli foundation. The work of JE was supported by the World Premier International Research Center Initiative (WPI), MEXT, Japan, and by the JSPS KAKENHI Grant Numbers 21H05451 and 21K20366. The work of GP is supported by grants from BSF-NSF,
Friedrich Wilhelm Bessel research award, GIF, ISF, Minerva, SABRA Yeda-Sela  WRC Program, the Estate of Emile Mimran, and the Maurice and Vivienne Wohl
Endowment.

\appendix

\section{Review of the relaxion mechanism}
\label{app:relaxion}

In this section we discuss the relaxation of the Higgs mass parameter. 
For the case of QCD relaxion, the back-reaction potential depends linearly on the Higgs vev as opposed to the quadratic case discussed in~\cite{Banerjee:2020kww}. 
A generic back-reaction potential which depends linearly on the Higgs vev can be written as 
\bea
V_{\rm br}= -\Lambda_{\rm b}^3 \left<H\right>\cos(\phi/f)\,,
\eea 
where $\Lambda_{\rm b}$ is the back-reaction scale. Following the notation of the main text Eq.~\eqref{eq:relaxion_potential}, the total relaxion potential can be written as
\begin{align}
V(\phi,H)&= \left(\Lambda^2-g\Lambda \phi\right)|H|^2+ \lambda |H|^4 \nn \\
		&-g\Lambda^3\phi  -\Lambda_{\rm b}^3 \left<H\right>\cos(\phi/f)\,.
\end{align} 
Below we set the Higgs quartic coupling $\lambda=1$ for notational convenience. 
We are interested in understanding the evolution of the relaxion close to the EW scale $(v)$ Higgs mass. 
In that case, the minimum of the potential can be found by solving two equations: $\partial V(\phi,H)/\partial |H|=0$ and $\partial V(\phi,H)/\partial \phi=0$.  
If $|\partial^2 V/\partial^2 H|\gg |\partial^2 V/\partial^2 \phi|$, then one can set the Higgs at its instantaneous minimum by solving $\partial V(\phi,H)/\partial |H|=0$.  
Using perturbation theory around the EW vacuum, one finds the relaxion-dependent Higgs vev as
\begin{align}
v^2(\theta_a) = \! \frac{v^2}{2}\!\left(\!-\frac{\Lambda^2}{v^2}\!+\frac{g\Lambda f \theta_a}{v^2} +\!\frac{\Lambda_{\rm b}^3}{2v^3}\cos\theta_a\!\!\right)\!\! +\mathcal{O}\!\left(\!\frac{\Lambda_{\rm b}^6}{v^6}\!\right)\!,\nn
\end{align}
where we write $\theta_a=\phi/f$. The perturbative expansion of the Higgs vev is valid as long as 
\bea
\Lambda_{\rm b}^3\ll v^3\,. 
\eea
From Eq.~\eqref{eq:QCD_axion_simplified} one can see that the above condition is easily satisfied for QCD axion. 
Setting the Higgs to its relaxion dependent vev, we obtain the effective potential of the relaxion as 
\bea
\!\!\!\!\!
V_{\rm eff}(\theta_a)= -g\Lambda^3 f\theta_a - (v^2(\theta_a))^2-\frac{\Lambda_{\rm b}^3v}{2}\cos\theta_a\,.
\eea
Thus, the relaxion encounters the first minimum when $V'_{\rm eff}(\theta_a)=0$ and we find,  
\bea
-g\Lambda^3 f - 2 v^2(\theta) v^2(\theta)'+\frac{\Lambda_{\rm b}^3v}{2}\sin\theta_a =0\,.
\label{eq:relaxion_stopping_1}
\eea 
By setting the EW scale as $g\Lambda^3f\simeq\Lambda_{\rm b}^3 v$, and defining a small parameter $\delta^2=\Lambda_{\rm b}^3/(v\Lambda^2)\ll1$, we find that the Higgs vev changes only incrementally as 
\bea 
\frac{\Delta v^2}{v^2} = \frac{v^2(\theta_a+ 2\pi)-v^2(\theta_a)}{v^2} = \pi \frac{g\Lambda f}{v^2}=\pi\delta^2 \,.
\eea
Following the calculation of~\cite{Banerjee:2020kww}, by realizing $\theta_a\to 2\pi m+ \theta_a$ where $m\in \mathbb{Z}$ and $\theta_a\in[0,2\pi)$ and then properly adjusting $m$ we find, 
\bea
\!\!\!\!\!\!\!
v_m^2(\theta_a) = v^2\left(1+m\pi\delta^2 + \frac{1}{2}\delta^2\theta_a+\frac{\Lambda_{\rm b}^3}{4v^3}\cos\theta_a\right).
\eea
From Eq.~\eqref{eq:relaxion_stopping_1} we get,
\bea
\frac{\sin\theta_a}{2}\left(1+\frac{v^2}{v_m^2(\theta_a)}\right)=\frac{v^2}{v_m^2(\theta_a)}+ \frac{v^2}{\Lambda^2}\,.
\eea
Note that, the effective potential written before and the above equation is valid only when the Higgs vev is close to $v$. By expanding $v_m(\theta_a)$ close to $v$ we find, the above equation admits a solution when,
\bea
\!\!\!\!\!\!\!
\sin\theta_a = 1-\frac{m\pi}{2}\delta^2 - \frac{1}{4}\delta^2\theta_a-\frac{\Lambda_{\rm b}^3}{8v^3}\cos\theta_a+\frac{v^2}{\Lambda^2}\,.\label{eq:sintheta}
\eea
It is easy to see that, the above equation has two solutions close to $\theta_a\sim\pi/2$.  
As the Higgs vev only increases incrementally with a small parameter $\delta$, we find the relaxion stopping point as
\bea
\theta_a-\pi/2\equiv\delta_\theta = \frac{\Lambda_{\rm b}^3}{8v^3}+\frac{\delta^2}{4} \mp \alpha \,\delta \,,
\eea
where $\alpha$ is some $\mathcal{O}(1)$ number. 
The mass of relaxion at the first minimum can be written as  
\bea
m_\phi^2 = \frac{\Lambda_{\rm b}^3 v}{f^2}\times \delta\,,
\eea
significantly reduced by the small parameter $\d$ compared to the naive expected value.

{\it Constraints:} For a successful relaxation of the Higgs mass we require the following conditions:
\bea
f\gtrsim\Lambda\gtrsim\Lambda_{\rm min}=4\pi v\,.
\label{eq:EFT_cons}
\eea
Here we are considering scanning of the Higgs mass during inflation. 
We require a separate inflaton sector dominates the energy of the universe during inflation and the classical evolution of the relaxion dominates over quantum spreading during inflation. 
These two requirements lead (respectively) to the constraints
\bea 
3H_I^2\Mpl^2\gtrsim \Lambda^4\,{\rm and}\,
(\Delta\phi)_{\rm cl}= \frac{g\Lambda^3}{3H_I^2}\gtrsim \frac{H_I}{2\pi}\,,
\label{eq:inf_cons}
\eea 
where $H_I$ is the Hubble scale during inflation. 

We also want the relaxion to be cosmologically stable in the first minima. This leads to the following constraint:
\bea
\frac{8\pi^2}{3} (g\Lambda^3 f)\, \delta^3\gtrsim H_I^4\,.
\label{eq:stab_cons}
\eea
In the case of a QCD relaxion, the back-reaction potential depends on the temperature and thus, it is only significant when $H_I<\Lambda_{\rm QCD}$. 
In this section we only consider inflationary based-relaxation of the Higgs mass with a back-reaction potential which depends linearly on Higgs vev. 
\\

Now let us consider the back-reaction potential of our interest as given in Eq.~\eqref{eq:QCD_backreaction}, \bea
V_{\rm br}(\phi) \sim &&-\Lambda_{\rm QCD}'^3  y'_u v'  \cos(\theta_a)\left(1 - \e_b \g\right) \nn \\ 
		&& - \Lambda_{\rm QCD}^3 \,y_u v\, \k \,\cos(N\theta_a)\,.
\eea 
Note that, as discussed in the relaxation is happening at the $k=0$ sector where all the quantities are denoted by $()'$. 
We see that in $V_{\rm br}$ the coefficient of $\cos\theta_a$ term is responsible for relaxion whereas the coefficient of $\cos(N\theta_a)$ term has contribution independent of the relaxing Higgs. 
To use the result of previous discussion, we can make the following replacements
\bea \label{eq:subs_normal_prime}
v &\to& v', \nn\\
\Lambda_b^3 &\to& \Lambda_{\rm QCD}'^3  y'_u \left(1 - \e_b \g\right), \\
V_{\rm eff}(\theta_a) &\to& V_{\rm eff}(\theta_a)-\Lambda_{\rm QCD}^3 \,y_u
v\, \k \,\cos(N\theta_a)\nn\,.
\eea
In the limit, $\Lambda_{\rm QCD}^3 \,y_u v\, \k\ll \Lambda_{\rm QCD}'^3  y'_u v' \left(1 - \e_b \g\right)$, using the above substitution, one obtains the relaxion stopping point as $\theta_a-\pi/2=\delta_\theta$ where 
\bea
\delta_\theta = \frac{y'_u\Lambda_{\rm QCD}'^3}{8v'^3}+\frac{N^2\k y_u\Lambda_{\rm QCD}^3 v}{y'_u \Lambda_{\rm QCD}'^3 v'} \mp \alpha\,\delta + \mathcal{O}(\delta^2) \,.
\eea
In the above equation we also use $\left(1 - \e_b \g\right)\simeq 1$. In the main text, for all the purposes we set $\alpha=1$. With the definition of $\epsilon_b=\Lambda_{\rm QCD}^3  y_u/(\Lambda_{\rm QCD}'^3  y'_u)$ and $\gamma=v/v'$, we get back Eq.~\eqref{eq:dtheta_QCD_relaxion}. 

Using the substitution \eqref{eq:subs_normal_prime}, we obtain the expression for relaxion mass
\bea
m_\phi^2 = \frac{y'_u\Lambda_{\rm QCD}'^3 v'}{f^2}\times \delta= \frac{y_u\Lambda_{\rm QCD}^3 v}{f^2}\times \frac{\delta}{\epsilon_b\g}\,.
\eea
All the constraints discussed before translate to this case with proper substitution given in Eq.~\eqref{eq:subs_normal_prime}. 
The additional constraint in this scenario comes from the fact that, as we consider both QCD and QCD' potential are temperature-dependent, we need 
\bea
H_I< \Lambda_{\rm QCD},\,\Lambda_{\rm QCD}'\,.
\label{eq:temp_cons}
\eea
Written explicitly, Eq.~\eqref{eq:EFT_cons}, becomes
\bea
f\gtrsim\Lambda\gtrsim 4\pi v'\,.
\eea
The form of Eq.~\eqref{eq:inf_cons} and \eqref{eq:stab_cons} do not change. However now one needs to replace
\bea
g\Lambda^3 f \simeq y_u' (\Lambda_{\rm QCD}')^3 v' = \frac{\L_a^4}{\e_b \g}\,.
\label{eq:hprime_vev}
\eea
Recall we define $\L_a = (\L_{\rm QCD}^3 m_u)^{1/4} = (\L_{\rm QCD}^3 y_u v)^{1/4}$ in the main text. Also, with the prime notation, 
\bea
\delta^2= \frac{y'_u (\Lambda_{\rm QCD}')^3}{v'\Lambda^2}=\frac{\Lambda_a^4}{v^2\Lambda^2}\frac{\g}{\e_b}\,.
\eea

In our parameter estimation, the constraints arising from a separate inflaton sector which dominates the energy of the universe during inflation Eq.~\eqref{eq:inf_cons} (left side), as well as stability of the first minimum Eq.~\eqref{eq:stab_cons}, were the most important. 
To estimate this constraint (the blue lines in Figure \ref{fig:HQQCD}), we used Eq.~\eqref{eq:hprime_vev} to fix $g\L^3 f$, and Eq.~\eqref{eq:inf_cons} (left side) to fix $H_I$; substituting both into Eq.~\eqref{eq:stab_cons} and solving for $\e_b$ recovers Eq.~\eqref{eq:conrel}.

It is straightforward to see that Eq.~\eqref{eq:inf_cons} (right side) and Eq.~\eqref{eq:temp_cons} are trivially satisfied. Observe from Eq.~\eqref{eq:stab_cons} that $H_I^4 \lesssim 8\pi \L_a^4\d^3/(3\e_b \g)$; this is at most $H_I \sim \keV$ for the largest $\d$ values we achieve, which are $\Ocal(10^{-10})$, and even if $\e_b\g \to 1$. Then Eq.~\eqref{eq:inf_cons} (right side) implies $H_I^3 \lesssim 2\pi\L_a^4/(3\e_b\g\,f)$, which is satisfied even if $f \to \Mpl$. Thus, in our case the requirement of classical evolution of the relaxion dominates over quantum spreading during inflation, provides a weaker constraint than the one provide by the cosmological stability of the relaxion.

\section{Full leading order and Higher harmonic corrections}
\label{app:QCD_prime}

In this section we investigate the back-reaction potential in greater detail. 
The back-reaction potential written in Eq.~\eqref{eq:QCD_backreaction} is
\bea
V_{\rm br}(\phi) \sim &&-\Lambda_{\rm QCD}'^3  y'_u v'  
\cos\left(\frac{\phi}{f}\right)\left(1 - \e_b \g\right) \nn \\ 
&& - \Lambda_{\rm QCD}^3 \,y_u v\, \k \,
\cos\left(\frac{N\phi}{f}\right)\,,
\label{}
\eea
which is only the leading order approximation of the full QCD-axion potential. Next we include the next-to-leading order corrections to the axion potential in the hidden sector (k=0 sector). 
We will consider below two possibilities for the relation between the up and down quark masses. 
The first is the isospin symmetric case, $y_u'=y_d'=y_q'$ and further below we shall comment on the case where $y_d = 2y'_u$ (both can be achieved in limit of flavor models). 

For the $y_u'=y_d'=y_q'$ case, including the NLO terms, the potential of the $k=0$ sector, can be written as 
\bea 
V_{\rm QCD'} &=& - 2 \Lambda_{\rm QCD}'^3\,  y'_q\, v'
	\left[ \,\cos \frac{\theta_a}{2}+\frac{y'_q\, v'}{\Lambda_{\rm QCD}'} \, R \cos\theta_a \right],\nonumber \\
	&=& -\Lambda_{\rm b}^3 v'\cos\frac{\th_a}{2} - \a^2\cos\th_a,
\eea   
where we define
\bea
\Lambda_b^3 = 2 y_q' \Lambda_{\rm QCD}'^3\,\,,\,\, \alpha^2 = 2 y_q'^2 \Lambda_{\rm QCD}'^2\, R\,,
\eea
and where $R\sim\mathcal{O}(10^{-3})$ denotes the NLO low energy efficient factors (see \cite{Lu:2020rhp} and refs. therein). 
The total back-reaction factor can be written as 
\bea
V_{\rm br}(\phi) = &&  V_{\rm QCD'} + \Lambda_{\rm QCD}^3 \,y_u v\, 
\sqrt{1+z^2+2 z\cos(\theta_a)}  
 \nn \\ 
&& \qquad \quad  - \Lambda_{\rm QCD}^3 \,y_u v\, \k \,
\cos\left(N\theta_{a}\right)\,.
\label{}
\eea
Note that the NLO correction to the SM QCD-axion potential is suppressed by a factor of order $\mathcal{O}\left(R\, m_{ u,d}/\Lambda_{\rm QCD}\right)\sim\mathcal{O}\left(10^{-5}-10^{-4}\right)$ compared to the LO SM term, and thus have been neglected here. 
As before, the full potential is obtained by adding in the scanning and Higgs-dependent potentials,
\begin{align}
 V_r = (\L^2 - g\L\phi)|H|^2 + |H|^4 - g\L^3 \phi\,,
\end{align}
where we set $\l=1$ as before.

To determine the relaxation of $v'$, one finds the relaxion-dependent vev using $\partial V/\partial |H| = 0$, which implies
\bea
v'^2(\theta_a) = \frac{v'^2}{2}\!\left(\!-\frac{\Lambda^2}{v'^2}\!+\frac{g\Lambda f \theta_a}{v'^2} 
+ \!\frac{\Lambda_{\rm b}^3}{2v'^3}\cos\frac{\theta_a}{2}
+\frac{\alpha^2}{v'^2}\cos\theta_a\!\!\right). \nn \\ 
\eea
The perturbative expansion is valid near the vev as long as 
\bea
\alpha \ll \Lambda_{\rm b} \ll v'.
\eea
Expanding the effective potential at $v'(\th_a)$, we obtain
\bea
V_{\rm eff}(\theta_a) && = -g\Lambda^3 f\theta_a - (v'^2(\theta_a))^2-\frac{\Lambda_{\rm b}^3v'}{2}\cos\frac{\theta_a}{2}
\nn\\
+&& \Lambda_{\rm QCD}^3 \,y_u v\left[\, 
\sqrt{1+z^2+2 z\cos(\theta)}  
 -  \k \,\cos\left(N\theta_a\right)\right]\,.   \nn
\eea

The first relaxion minima can be found when $V'_{\rm eff}(\theta_a)=0$, and can be written as
\bea
&&\frac{\sin\theta_a/2}{2}\left[1+\frac{v'^2}{v'^2(\theta_a)}\right]  
= \frac{v'^2}{\Lambda^2}+ \frac{2\alpha^2 v'}{\Lambda_{\rm b}^3}\sin\theta_a \nn \\
&& + \frac{v'^2}{v'^2(\theta_a)}\left[1-N\k\e_b\g\cos(N\theta_a)+\frac{\e_b \g z\cos\theta_a}{\sqrt{1+z^2+2 z\cos(\theta)}} \right], 
\nn \\
\, \nn 
\eea  
where $\e_b,\g\ll 1$ were defined in Eq.~\eqref{eq:breaking}.
 
Expanding near the vev, it has been shown that one can write \cite{Banerjee:2020kww}
\bea
\frac{v'^2(\theta_a)}{v'^2} \simeq \left(1+\beta\delta^2 +\frac{\delta^2}{4}\theta_a+ \!\frac{\Lambda_{\rm b}^3}{4v'^3}\cos\frac{\theta_a}{2}-\frac{\alpha^2}{2 v'^2}\cos\theta_a\!\!\right)\nn
\eea
where we neglected terms of the order of $\mathcal{O}(\alpha^2\Lambda_{\rm b}^3/v'^5)$ and $\beta\in(0,1)$ denotes some $\mathcal{O}(1)$ number. 

Plugging everything back, we get
\bea
\sin\frac{\theta_a}{2} &\simeq& 1 -\frac{\beta}{2}\delta^2 -\frac{\delta^2}{8}\theta_a- \!\frac{\Lambda_{\rm b}^3}{8v'^3}\cos\frac{\theta_a}{2}+\frac{\alpha^2}{2 v'^2}\cos\theta_a\!\!\nn \\
&& - N\k\e_b\g\sin(N\theta_a)+ \frac{\e_b\g z\sin\theta_a}{\sqrt{1+z^2+2 z\cos\theta_a}}  \nn \\ 
&& + \frac{v'^2}{\Lambda^2} + \frac{2\alpha^2 v'}{\Lambda_{\rm b}^3}\sin\theta_a\,. \nn
\eea 
The above equation admits a solution when $\theta_a=\pi-\delta_\theta$
with 
\bea
\delta_\theta \simeq \frac{\Lambda_b^3}{4v'^3}+N^2\k\e_b\g  \mp \beta\,\delta
		+ \frac{z}{1-z}\e_b\g+\frac{2\alpha^2 v'}{\Lambda_{\rm b}^3}\,. 
\eea

The upshot is, we need 
\bea
\e^2 = \frac{y_u'\L_{\rm QCD}'^3}{v'^3} &\lesssim& \theta_{\rm CP}\,,\\
N^2\k\e_b\g = \frac{N^2\kappa y_u\Lambda_{\rm QCD}^3 v}{y_q'\Lambda_{\rm QCD}'^3 v'}&\lesssim & \theta_{\rm CP}\,, \label{eq:appkappa}\\
 \d^2 = \frac{y_q' \Lambda_{\rm QCD}'^3}{v'\Lambda^2}&\lesssim & \theta_{\rm CP}^2\,, \label{eq:appdelta} \\
 \frac{z}{1-z}\e_b \g = \frac{z}{1-z}\frac{y_u\Lambda_{\rm QCD}^3 v}{y_q'\Lambda_{\rm QCD}'^3 v'}&\lesssim & \theta_{\rm CP}\,,\label{eq:appzz}\\
 R\frac{m_q'}{\Lambda_{\rm QCD}'}&\lesssim & \theta_{\rm CP}\,. \label{eq:appmqlam}
\eea 
Observe that the first three constraints are similar to the ones discussed for the toy model Eqs.~(\ref{eq:epsqu}-\ref{eq:kappaconstraint}) coming from the relaxation requirement and the $\mathbb{Z}_{N}$ symmetric part of the backreaction potential.  
As discussed before, both the SM QCD potential and the NLO contributions of the ${\rm QCD}'$ potential shift the relaxion stopping point and these two constraints are shown by Eq.~\eqref{eq:appzz} and Eq.~\eqref{eq:appmqlam} respectively. 
Note also that Eq.~\eqref{eq:appkappa} provides a weaker constraint than Eq.~\eqref{eq:appzz} for any $N\geq 2$.

Saturating Eq.~\eqref{eq:appzz} and substituting into Eq.~\eqref{eq:appdelta} gives
\bea \label{eq:Lam1}
\Lambda \gtrsim \left(\frac{z}{1-z}\frac{1}{\theta_{\rm CP}^3} \frac{\Lambda_a^4}{v'^{2}}\right)^{1/2}.
\eea 
Saturating \ref{eq:appzz} and substituting this into the relaxion constraint of \ref{eq:conrel} gives
\bea \label{eq:Lam2}
\!\!\!\!\!\!\!\!\!\!\!\!\!
\L &\lesssim & \left[24\pi^2 \left(\frac{z}{(1-z)\theta_{\rm CP}}\right)^{5/2}\left(\frac{\L_a^{10} \Mpl^4}{v'^3}\right)\right]^{1/11}\!.
\eea
Combining Eq.~\eqref{eq:Lam1}  and Eq.~\eqref{eq:Lam2}  gives 
\bea \label{eq:vpcon}
v'\gtrsim \left[\left(\frac{z}{1-z}\right)^3\frac{1}{\theta_{\rm CP}^{14}}\frac{\L_a^{12} }{24\pi^2\,\Mpl^4} \right]^{1/8}.
\eea

All the constraints can be satisfied for 
\bea
\theta_{\rm CP} = 10^{-9}, 
\,\frac{v}{v'} = 4\times 10^{-3},
\,\frac{\Lambda_{\rm QCD}}{\Lambda_{\rm QCD}'} = 7\times 10^{-4}\,,   \nn
\eea
where we saturated the inequalities of Eqs.~(\ref{eq:vpcon}-\ref{eq:LQCDcon}).
If we vary $y_q'v'\Lambda_{\rm QCD}'^3$, then we can  get $v'\sim 20\TeV$. 
\\

An even smaller $v'\sim \TeV-10\TeV$, can be achieved if the $\rm{QCD}'$ sector has some flavour symmetry which protects the ratio of the hidden quark masses to be $1/2$.    
In this case, by repeating the same procedure as above, one obtains the stopping point be close to $2\pi/3$ in the hidden sector. 
The LO SM correction to the stopping point can be calculated as for $\theta_a =2\pi/3-\delta_\theta$  
\bea
\!\!\!\!\!
\frac{\epsilon_{b}\gamma z\sin\theta_a}{\sqrt{1+z^2+2 z\cos\theta_a}}
		&\simeq& \epsilon_{b}\gamma\, \delta_\theta \, \frac{z(2-5 z+2 z^2)}{4(1-z+z^2)^{3/2}}\nn\\
		&=& \epsilon_{b}\gamma\, \delta_\theta \begin{cases} 8\times 10^{-3}  \,\,\,\,\,\,\,{\rm for}\,\, z=0.485\\
2\times 10^{-3}  \,\,\,\,\,\,\,{\rm for}\,\, z=0.504\\	 
1.9\times 10^{-2}  \,\,{\rm for}\,\, z=0.466\\
	\end{cases}	\nn\\
\eea 
where $z$ is the SM ratio of the up and the down quark masses. 
The values above are the central value ($z=0.485$) and $2\s$ ($z=0.504,0.466$) results reported in~\cite{Fodor:2016bgu}.  
Then the constraint Eq.~\eqref{eq:appzz} changes to
\bea
\!\!\!\!\!\!\!\!\!\!\!\!\!
\frac{z}{1-z}\frac{y_u\Lambda_{\rm QCD}^3 v}{y_q'\Lambda_{\rm QCD}'^3 v'}\lesssim  \theta_{\rm CP} \to 2 C_z \frac{y_u\Lambda_{\rm QCD}^3 v}{y_q'\Lambda_{\rm QCD}'^3 v'}\lesssim  \theta_{\rm CP}\,, 
\eea 
with 
$C_z=(0.008,0.002,0.019)$. As such, one can use the substitution $z/(1-z) \to 2C_z$ in the results above to determine the parameter range relevant for this case.

By saturating Eq.~\eqref{eq:appmqlam} and Eq.~\eqref{eq:conrel} we find that all the constraints are satisfied for 
\bea
&\theta_{\rm CP}=& 10^{-9},  \,\,\,z = 0.485,\,\,\, v'=9 \TeV\,\\
&\theta_{\rm CP}=& 10^{-9},  \,\,\,z = 0.504,\,\,\, v'=6 \TeV\,\nonumber\\
&\theta_{\rm CP}=& 10^{-9},  \,\,\,z = 0.466,\,\,\, v'=13 \TeV\,.\nonumber
\eea

\section{Generating large hierarchy between the confinement scales of the hidden and visible sectors}
\label{app:large_alphas}

In this section we discuss how to generate a large hierarchy between the confinement scales of the hidden sector and the SM. 
To achieve this, we add additional $\Delta N_f$ number of heavy vector-like fermions with mass $m_{\rm NP}'$ and $m_{\rm NP}$ in the $k=0$ sector and the SM respectively. Adding additional fermions with different masses requires breaking of the $\mathbb Z_{N}$ symmetry. 
Thus, in order for the $\mathbb Z_{N}$ symmetry to be realized for the axion, we require \bea
m_{\rm NP}'\,,\, m_{\rm NP} \lesssim f\,,
\eea
where $f$ is the Peccei-Quinn symmetry breaking scale. 

Let us first consider the $k=0$ (hidden) sector. With the additional vector-like fermions whose mass is larger than the hidden top quark mass, $m_{t}'$, we consider the running of 
$\alpha_s$. 
For the energy scale $\mu$, $m_{t'}\lesssim \mu\lesssim m_{\rm NP}'$, we can write at the 1-loop order, 
\bea
\frac{2\pi}{\alpha_s(m_t'+\epsilon)} &=& \frac{2\pi}{\alpha_s(f)}
- b(n_f+\Delta n_f) \ln\frac{f}{m_t'}\nn\\ 
&&-\frac{2\Delta n_f}{3}\ln\frac{m_{\rm NP}'}{m_t'}\,, \nn
\eea
with $b(n)=11-2n/3$. Now for a scale $\Lambda>  m_u',m_d'$, we can write   
\bea 
\frac{2\pi}{\alpha_s(\Lambda)} &=& \frac{2\pi}{\alpha_s(m_t')}
- b(n_f)\ln\frac{m_t'}{\Lambda}-\frac{8}{3}\ln\frac{v'}{\Lambda} \nn\\
&=& \frac{2\pi}{\alpha_s(f)}
- b(n_f+\Delta n_f) \ln\frac{f}{m_t'}-\frac{2\Delta n_f}{3}\ln\frac{m_{\rm NP}'}{m_t'}\nn\\
&-&  b(n_f)\ln\frac{m_t}{\Lambda}-\frac{8}{3}\ln\frac{v'}{\Lambda}\,. 
\eea
Thus, one obtains the confinement scale, $\Lambda'_{\rm IR}$ when $\alpha_s^{-1}(\Lambda'_{\rm IR})\to 0$ as, 
\bea
\Lambda_{\rm IR}'\sim f \left(\frac{v'}{f}\right)^{\alpha_1}\left(\frac{m_{\rm NP}'}{f}\right)^{\alpha_2}\exp[-2\pi/\alpha_s(f)]\,,
\eea 
with 
\bea
\!\!\!\!\!\!
\alpha_1= \frac{\frac{8}{3}}{b(6)+\frac{8}{3}}\simeq 0.28\,,\,\, \alpha_2=\frac{\frac{2\Delta n_f}{3}}{b(6)+\frac{8}{3}}=\frac{2\Delta n_f}{29}\,.
\eea
We repeat the same exercise for the SM sector and obtain the same result with $v'\to v$ and $m'_{\rm NP}\to m_{\rm NP}$. We also assume $m_{\rm NP}>m_t$ where $m_t\sim 175 \GeV$ is the SM top quark mass. 
So the ratio of the confinement scale can be obtained as 
\bea
\frac{\Lambda_{\rm QCD}'}{\Lambda_{\rm QCD}}\sim \frac{\Lambda_{\rm IR}'}{\Lambda_{\rm IR}}=\left(\frac{v'}{v}\right)^{0.28}\left(\frac{m_{\rm NP}'}{m_{\rm NP}}\right)^{2\Delta n_f/29}\,, 
\eea 
where $\Delta n_f$ is the number of extra vector like fermions that we are adding. 
For, $\Delta n_f=4$, $m_{\rm NP}/m_{\rm NP}'=10^{-9}$, $v/v'=10^{-3}$ one obtains   
\bea
\frac{\Lambda_{\rm QCD}}{\Lambda_{\rm QCD}'}\sim \frac{\Lambda_{\rm IR}}{\Lambda_{\rm IR}'}=4.8 \times 10^{-4}\,.  
\eea
The current bound on the vector-like fermion mass scale $m_{\rm NP}$ is at the level of $1-2$ TeV, though this is somewhat model-dependent (see e.g.~\cite{ATLAS:2018ziw}).

\bibliographystyle{apsrev4-1}
\bibliography{quality.bib}

\end{document}